\documentclass[10pt,twocolumn]{article}

\usepackage[utf8]{inputenc} 
\usepackage[T1]{fontenc}    
\usepackage{url}            
\usepackage{booktabs}       
\usepackage{amsfonts}       
\usepackage{nicefrac}       
\usepackage{microtype}      

\usepackage{cvpr}

\usepackage{times}
\usepackage{epsfig}
\usepackage{graphicx}
\usepackage{amsmath}
\usepackage{amssymb}
\usepackage{paralist}
\usepackage{dsfont}
\usepackage{multirow}
\usepackage{wrapfig}
\usepackage{xcolor}
\usepackage{subfigure}
\usepackage{pifont}
\usepackage{lipsum}
\input{taesup.sty}

\newcommand{\cmark}{\ding{51}}%
\newcommand{\xmark}{\ding{55}}%

\newcommand{\tsmcolor}{black}
\usepackage{pdfpages}

\usepackage{footnote}
\usepackage{fixltx2e}

\usepackage{tabularx,booktabs}
\newcommand{\multiline}[1]{%
  \begin{tabularx}{\dimexpr\linewidth-\ALG@thistlm}[t]{@{}X@{}}
    #1
  \end{tabularx}
}

\usepackage[pagebackref=true,breaklinks=true,letterpaper=true,colorlinks,bookmarks=false]{hyperref}
 \cvprfinalcopy 
\def\cvprPaperID{5040} 

\ifcvprfinal\fi
\begin{document}

\title{FBI-Denoiser: Fast Blind Image Denoiser for Poisson-Gaussian Noise}


\author{Jaeseok Byun\textsuperscript{\rm 1}\thanks{Equal contribution.}, Sungmin Cha\textsuperscript{\rm 1}\footnotemark[1], ~and Taesup Moon\textsuperscript{\rm 2}\thanks{Corresponding author (E-mail: \texttt{tsmoon@snu.ac.kr})} \\\
\textsuperscript{\rm 1}Department of Electrical and Computer Engineering, Sungkyunkwan University, Suwon, Korea\\
\textsuperscript{\rm 2}Department of Electrical and Computer Engineering, Seoul National University, Seoul, Korea\\
{\tt\small \{wotjr3868, csm9593\}@skku.edu,
\tt\small tsmoon@snu.ac.kr}
}
\maketitle
\thispagestyle{empty}
\begin{abstract}
We consider the challenging blind denoising problem for Poisson-Gaussian noise, in which no additional information about clean images or noise level parameters is available. Particularly, when only ``single'' noisy images are available for training a denoiser, the denoising performance of existing methods was not satisfactory. 
Recently, the blind pixelwise affine image denoiser (BP-AIDE) was proposed and significantly improved the performance in the above setting, to the extent that it is competitive with denoisers which utilized additional information. However, BP-AIDE seriously suffered from slow inference time due to the inefficiency of noise level estimation procedure and that of the blind-spot network (BSN) architecture it used. 
To that end, we propose Fast Blind Image Denoiser (FBI-Denoiser) for Poisson-Gaussian noise, which consists of two neural network models; 1) PGE-Net that estimates Poisson-Gaussian noise parameters $2000$ times faster than the conventional methods and 2) FBI-Net that realizes a much more efficient BSN for pixelwise affine denoiser in terms of the number of parameters and inference speed.
Consequently, we show that our FBI-Denoiser blindly trained solely based on single noisy images can achieve the state-of-the-art performance on several real-world noisy image benchmark datasets with much faster inference time ($\times 10$), compared to BP-AIDE. The official code of our method is available at \href{https://github.com/csm9493/FBI-Denoiser}{https://github.com/csm9493/FBI-Denoiser}.
 
   
\end{abstract}








\section{Introduction}

\textit{Convolutional neural network} (CNN)-based denoisers achieved impressive state-of-the-art denoising performances mainly by utilizing the supervised learning approach based on collecting many clean and noisy image pairs. The performance gain was first shown in the additive white Gaussian noise setting \cite{(DnCNN)zhang2017beyond, (Memnet)TaiYanLiuXu17, (RNAN)zhang2019residual, (MWCNN)liu2018multi, (N3Net)plotz2018neural, (SADNet)chang2020spatial}, then the approach was extended also to the  Poisson-Gaussian noise setting, which better models the real-world source-dependent noise. It was demonstrated that the gain was attained not only with respect to the quantitative metrics, \textit{e.g.}, PSNR or SSIM \cite{(SSIM)wang2004image}, on the real-world noise benchmarks such as DND \cite{(DND)plotz2017benchmarking}, SIDD \cite{(SIDD)SIDD_2018_CVPR} and FMD \cite{(FMD)zhang2019poisson}, but also with respect to the inference time for denoising (using GPUs), compared to the more conventional prior- or optimization-based methods \cite{(BM3D)dabov2007image,(WNNM)gu2014weighted}. 

Despite above promising achievements, the plain supervised learning approach has a critical drawback in a more practical, real-world setting, since the availability of the enough number of the clean-noisy image pairs for training is sometimes a luxury that cannot be simply assumed. For example, in medical imaging (CT or MRI), obtaining the underlying clean image for a noisy image becomes extremely time-consuming and expensive. In order to overcome this drawback, several attempts have been made in recent years. 
The first approach is to utilize \textit{unpaired} clean images and generate synthetic noisy images to again carry on the supervised training with the generated pairs. For example, in \cite{(UPI)brooks2019unprocessing, (CycleISP)zamir2020cycleisp}, based on in-camera signal processing (ISP) pipeline and specific Poisson-Gaussian noise parameters, they generated synthetic noisy sRGB or rawRGB images from clean sRGB images. Another examples can be found in \cite{(Unpaired)wu2020unpaired, (GCBD)chen2018image}, in which they learned a model to \textit{generate} noise present in the given noisy images, then used that model to corrupt the clean images to build paired supervised training set. While these approaches were shown to achieve good performance for some specific settings, they either lack generalities or have limited performance for real-world noisy image denoising.
The second recent approach to remove the requirement of clean-noisy pairs is to train a denoiser solely based on noisy images \cite{(N2N)lehtinen2018noise2noise, (N2V)krull2018noise2void, (N2S)batson2019noise2self, (HQDenoising)laine2019high, (Self2Self)Quan_2020_CVPR, (Noiser2Noise)moran2020noisier2noise, (GAN2GAN)chagan2gan, (Sure)soltanayev2018training}. However, those methods also had their own limitations, such as requiring pairs of independently realized noisy images for the same clean source \cite{(N2N)lehtinen2018noise2noise, (eSure)zhussip2019extending}, poor performance on benchmark datasets \cite{(N2V)krull2018noise2void,(N2S)batson2019noise2self}, large inference time due to requiring many number of samplings \cite{(Self2Self)Quan_2020_CVPR}, 
and limited or no experiment on real-world noise setting \cite{(HQDenoising)laine2019high, (Noiser2Noise)moran2020noisier2noise, (Sure)soltanayev2018training, (eSure)zhussip2019extending, (GAN2GAN)chagan2gan}.

Recently, BP-AIDE \cite{(BPAIDE)9117146}, which extends the framework of \cite{(FCAIDE)cha2019fully,(NAIDE)cha2018neural}, was proposed as another attempt to lift the requirement of clean images. Namely, the scheme made a unique combination of Generalized Anscombe Transformation (GAT) \cite{(GAT)anscombe1948transformation}, Poisson-Gaussian noise estimation \cite{(pg_Liu)liu2014practical, (Foi)foi2008practical}, and an unbiased estimator of MSE for pixelwise affine denoisers proposed in \cite{(NAIDE)cha2018neural} in order to train a \textit{blind} denoiser for Poisson-Gaussian noise solely on single noisy images with no additional information. The method achieved the state-of-the-art performance on several benchmarks for real-world noise \cite{(FMD)zhang2019poisson,(SIDD)SIDD_2018_CVPR,(DND)plotz2017benchmarking} compared to the methods \cite{(N2V)krull2018noise2void,(BM3D)dabov2007image} that operate in the identical setting. However, BP-AIDE had one critical limitation; it suffers from the slow inference time due to the following two reasons. First, for every given noisy image, BP-AIDE must separately carry out the Poisson-Gaussian noise parameter estimation, which usually takes a couple of seconds for moderately sized images. Second, the method simply utilizes the so-called blind-spot network (BSN) architecture proposed in \cite{(FCAIDE)cha2019fully} as a denoiser, but the corresponding structure is quite complex and requires large GPU memory, leading to a slow inference time. 


To tackle above limitation, we make two significant improvements on BP-AIDE and propose Fast Blind Image  Denoiser (FBI-Denoiser). 
 Firstly, we propose PGE (Poisson-Gaussian Estimation)-Net, which \textit{learns} to estimate the Poisson-Gaussian noise parameters solely from noisy images by converting GAT and Gaussian noise estimation steps to tensor operations and by proposing a novel loss function. 
 Secondly, we devise FBI-Net, a new compact fully convolutional BSN, which performs almost the same as the network in \cite{(FCAIDE)cha2019fully} but significantly reduces the inference time. The FBI-Denoiser is trained in two steps; first train PGE-Net with noisy images, then train FBI-Net again with the same noisy images and the outputs of PGE-Net, following the procedures of BP-AIDE. As a result, we significantly improve the inference time of BP-AIDE ($\times 10$ speed-up) as well as achieve the state-of-the-art \textit{blind} denoising performance on real-world noise benchmarks \cite{(FMD)zhang2019poisson,(SIDD)SIDD_2018_CVPR,(DND)plotz2017benchmarking}. 
 

\section{Related Works}

\noindent\textbf{Neural network based blind image denoising} 
\textcolor{\tsmcolor}{As mentioned above, several \textit{blind} image denoisers are proposed to resolve the issue of the dependence on clean images for training. Table \ref{table:related_work_summary} summarizes and compares the settings of recently proposed schemes.
A variety of schemes \cite{(HQDenoising)laine2019high,(Sure)soltanayev2018training, metzler2018unsupervised,(Noiser2Noise)moran2020noisier2noise,(GAN2GAN)chagan2gan,(eSure)zhussip2019extending} were proposed, but their applicability to the Poisson-Gaussian noise was limited.
Noise2Noise (N2N) \cite{(N2N)lehtinen2018noise2noise} has been shown to be effective for Poisson-Gaussian noise setting, but 
it still required \textit{two} independent realizations of noisy images for the same source, which is not practical.
For address such limitation of N2N, Noise2void (N2V) \cite{(N2V)krull2018noise2void}, Noise2self (N2S)\cite{(N2S)batson2019noise2self}, and BP-AIDE \cite{(BPAIDE)9117146} adopted self-supervised learning approach which can be solely trained with single images corrupted by Poisson-Gaussian noise. Their settings fully coincide with ours, but they suffer from either poor performance or slow inference time. 
More recently, D-BSN \cite{(Unpaired)wu2020unpaired} was proposed for the setting with unpaired clean and noisy images.
It also contains self-supervised (self-sup) learning step that improved N2V by elaborating the blind spot network architecture and pixelwise noise level estimation network, but we show that our FBI-Denoiser significantly outperforms it.}

\begin{table}[h!]\vspace{-.0in}\caption{Summary of different settings among the recent baselines.}
\centering
\smallskip\noindent
\resizebox{.98\linewidth}{!}{
\begin{tabular}{|c||c|c|}
\hline
Alg. \textbackslash  Requirements                               & Noisy pairs & Poisson-Gaussian noise \\ \hline \hline
\begin{tabular}[c]{@{}c@{}}HQ-SSL \cite{(HQDenoising)laine2019high}, SURE \cite{(Sure)soltanayev2018training, metzler2018unsupervised}, \\ Noiser2Noise \cite{(Noiser2Noise)moran2020noisier2noise}, GAN2GAN \cite{(GAN2GAN)chagan2gan}\end{tabular}         & \xmark           & \xmark                      \\ \hline
Ext. SURE \cite{(eSure)zhussip2019extending}                                                                           & \cmark           & \xmark                      \\ \hline
N2N \cite{(N2N)lehtinen2018noise2noise}                                                                                 & \cmark           & \cmark                      \\ \hline \hline
\begin{tabular}[c]{@{}c@{}}N2V \cite{(N2V)krull2018noise2void}, N2S \cite{(N2S)batson2019noise2self}, BP-AIDE \cite{(BPAIDE)9117146}, \\ D-BSN \cite{(Unpaired)wu2020unpaired} (Self-Sup), Self2Self \cite{(Self2Self)Quan_2020_CVPR}\end{tabular}  & \xmark           & \cmark                      \\ \hline
FBI-Denoiser                                                                         & \xmark           & \cmark                      \\ \hline
\end{tabular}}
\label{table:related_work_summary}
\end{table}


\noindent\textbf{Traditional denoising method} \ \ The classical denoising methods, \eg, wavelet-based \cite{(wavelet-based-1)donoho1995adapting}, filtering-based \cite{(filtering-based-1)buades2005review, (BM3D)dabov2007image}, optimization-based \cite{(optimization-based-1)elad2006image,(optimization-based-2)mairal2009non,(WNNM)gu2014weighted} and effective prior-based \cite{(effective-prior-based-1)zoran2011learning}, are typically capable of denoising with only single noisy images. However, since the training procedure with multiple images is absent in these methods, they suffer from large inference time and limited performance. 

\noindent\textbf{Noise estimation method}\ 
Most of above methods assume that prior knowledge about noise characteristics is given, but, it is typically unavailable in practice. To alleviate this unrealistic assumption, several noise estimation methods have been proposed, especially for two well-known noise models: additive white Gaussian noise (AWGN) and Poisson-Gaussian noise model. For AWGN, the noise variance of an image is assumed to be constant over all pixel values, \textit{i.e.}, the only parameter is the noise variance. 
Recently, low-rank patch selection methods \cite{(gaussian_liu)liu2013single, (gaussian_pyatykh)pyatykh2012image} using principal component analysis (PCA) showed the state-of-the-art performance in the AWGN case. \cite{(chen)chen2015efficient} further refined this approach by resolving the underestimation problem of \cite{(gaussian_liu)liu2013single, (gaussian_pyatykh)pyatykh2012image} through statistical analysis of the eigenvalues. 
 Unlike the case of AWGN, the Poisson-Gaussian noise model \cite{(Foi)foi2008practical}, which is often used to characterize the real source-dependent noise in the raw-sensed images, has a heterogeneous noise variance and two parameters $(\alpha, \sigma)$. Most existing methods \cite{(Foi)foi2008practical,(pg_abramov)abramov2010improved, (pg_uss)uss2013image,(pg_Liu)liu2014practical} for estimating Poisson-Gaussian noise first obtain the local estimated means and variances, then fit the noise model with these local estimates using maximum likelihood estimation (MLE).
 \cite{(Foi)foi2008practical} first proposed the Poisson-Gaussian noise model and an estimation algorithm for it using wavelet decomposition. Recently, \cite{(pg_Liu)liu2014practical} extended this approach to the generalized source-dependent noise by suggesting iterative patch selection method. 
 
\section{Problem Setting and Preliminaries}

\subsection{Notations}
We denote $\bm x\in\mathbb{R}^n$ as the clean image and $\bm Y \in\mathbb{R}^n$ as its noise-corrupted observation. The real-world image sensor noise is generally modeled with the Poisson-Gaussian noise model \cite{(Foi)foi2008practical} which consists of two mutually independent components. Under this model, the $i$-th noisy pixel becomes 
 \begin{align}
 Y_{i} = \alpha P_i + N_{i}, \ \  i=1,\dots,n \label{eq:noise model}
 \end{align}
 in which $P_{i} \sim \text{Poisson}({x}_{i})$ is the source-dependent Poisson noise with mean $x_i$ caused by photon sensing, $N_{i} \sim \mathcal{N}(0, \sigma^2)$ is the remaining signal-independent Gaussian noise. Here, $\alpha > 0$ is a scaling factor which depends on the sensor and analog gain, and $\sigma$ is the standard deviation of the Gaussian noise. For each clean and noisy image, we assume that each pixel is normalized and clipped to have a value in the range $[0,1]$.
Thus, the Poisson-Gaussian noise model is characterized by the parameters $(\alpha, \sigma)$ and the noise variance of $\bm Y$ can be represented as
\begin{equation}
\text{ Var} \,[\bm Y\,|\,\bm x] = \alpha^2\bm x+\sigma^2 \label{eq:noise_variance}
\end{equation}

\subsection{Generalized Anscombe Transformation (GAT)}\label{sec:gat}
A common approach for denoising the noisy image corrupted with Poisson-Gaussian noise is to apply GAT \cite{(GAT)anscombe1948transformation}, which transforms each pixel $Y_i>0$ to  
\begin{equation}
G_{\alpha,\sigma}(Y_{i}) = 
 \frac{2}{\alpha} \sqrt{\alpha Y_i + \frac{3}{8} {\alpha}^{2}+{\sigma}^2}. 
\label{eq:gat}
\end{equation}
The transform (\ref{eq:gat}) is known to stabilize the noise of each pixel of the transformed image, $\bm G_{\alpha,\sigma}(\bm Y)$, such that it approximately becomes Gaussian with unit variance. Using this property, we can have a simple denoising scheme for Poisson-Gaussian noise. Namely, a general Gaussian denoiser can be applied to $\bm G_{\alpha,\sigma}(\bm Y)$ to obtain a denoised version $\bm D$.
Then, by denoting $D_i$ as the $i$-th pixel of $\bm D$, the estimate of the original clean image $\bm x$ is obtained by applying the Inverse Anscombe transformation (IAT), which is often approximated by \cite{(IAT)makitalo2012optimal}
\begin{align}
&I_{\alpha,\sigma}({D}_{i})\label{eq:IAT} \\
= &\frac{1}{4}{D}^{2}_{i}+\frac{1}{4}\sqrt{\frac{3}{2}}{D}_{i}^{-1} -\frac{11}{8}{D}_{i}^{-2} 
+  \frac{5}{8}\sqrt{\frac{3}{2}}{D}^{-3}_{i}-\frac{1}{8}-\frac{{\sigma}^{2}}{{\alpha}^{2}}\nonumber
\end{align}
Thus, the final denoised image becomes $\hat{\bm x}=\bm I(\bm D)$. Note both the GAT and IAT require noise parameters ($\alpha$, $\sigma$). 
\subsection{BP-AIDE} \label{subsec:bp-aide}
As briefly mentioned in the Introduction, BP-AIDE \cite{(BPAIDE)9117146} combined GAT (\ref{eq:gat}) and Poisson-Gaussian noise estimation methods \cite{(pg_Liu)liu2014practical, (Foi)foi2008practical}, which estimates $\alpha$ and $\sigma^2$ of (\ref{eq:noise model}), with a pixelwise affine denoiser developed in \cite{(FCAIDE)cha2019fully}. The method consists of 3 steps, (1) Est+GAT (2) Training (3) Inference, and we briefly review each step below.



\noindent\textbf{(1) Est+GAT}\ Use the Poisson-Gaussian noise estimation methods \cite{(pg_Liu)liu2014practical, (Foi)foi2008practical} to obtain the estimated noise parameters $(\hat{\alpha},\hat{\sigma})$. Then, apply GAT  (\ref{eq:gat}) with the estimated parameters and obtain the transformed image $\bm G_{\hat{\alpha},\hat{\sigma}}(\bm Y)$. Then, the normalized version of the transformed image is obtained by $\bm Z\triangleq \bm (G_{\hat{\alpha},\hat{\sigma}}(\bm Y)-m)/\beta$, in which $m=\min_i G_{\hat{\alpha},\hat{\sigma}}(Y_i)$ and $\beta=\max_i G_{\hat{\alpha},\hat{\sigma}}(Y_i)-\min_i G_{\hat{\alpha},\hat{\sigma}}(Y_i)$. Note each pixel in $\bm Z$
has a value in $[0,1]$, and the variance of the noise becomes approximately $\beta^{-2}$.



\begin{figure*}[h]
    \centering
    \subfigure{
    \includegraphics[width=0.8\linewidth]{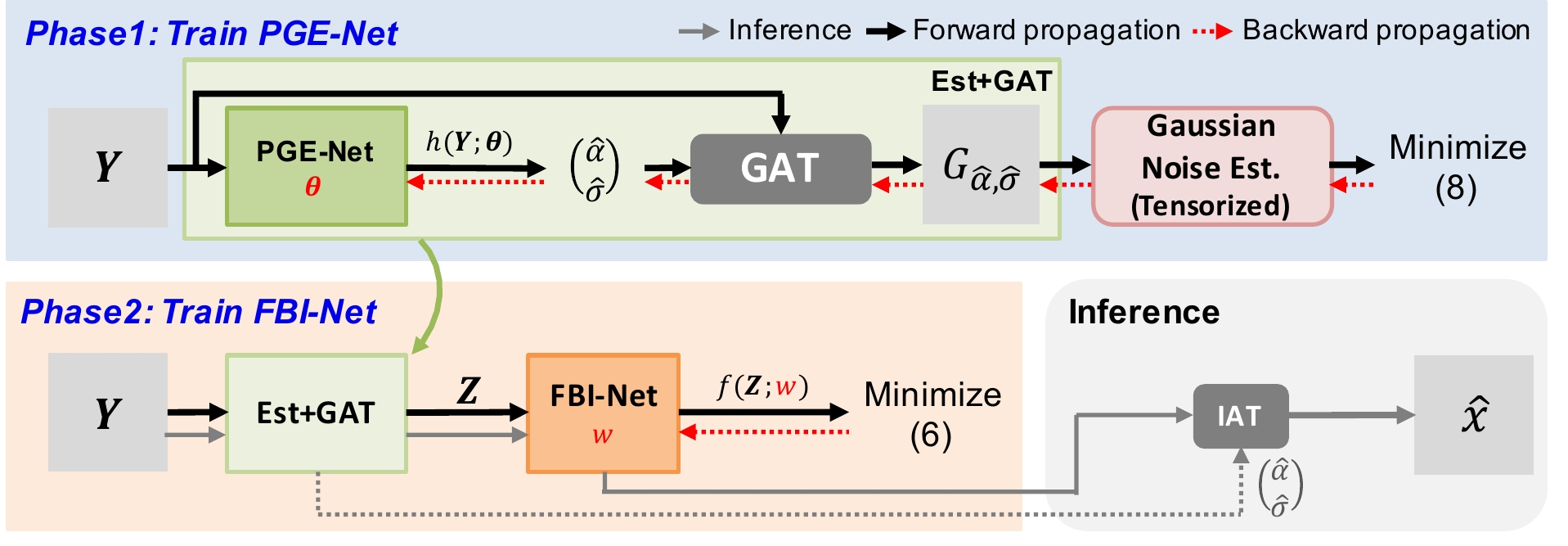}}
    \caption{Overall process of FBI-Denoiser.}
    \label{fig:overall_process}
    \vspace{-.2in}
\end{figure*}
\noindent\textbf{(2) Training}\ Given the normalized, transformed noisy image $\bm Z$, BP-AIDE trains a pixelwise affine denoiser $\bm f(\bm Z;\wb)\in\mathbb{R}^n$, proposed in \cite{(FCAIDE)cha2019fully}, of which $i$-th reconstruction is 
\begin{align}
f_{i}(\bm{Z};\wb)= a_{1}(\wb, \Cb_{k\times k}^{-i})\cdot Z_{i} + a_{0}(\wb, \Cb_{k\times k}^{-i}). 
\label{eq:affine}
 \end{align}
 In (\ref{eq:affine}), $\Cb_{k\times k}^{-i}$ denotes the $k\times k$ noisy patch surrounding ${Z}_i$ that \textit{excludes} $Z_i$, and $\{a_m(\wb,\Cb_{k\times k}^{-i})\}_{m=0,1}$ are the outputs of a specially designed fully convolutional network with parameter $\wb$ that takes $\bm Z$ as input, but guarantees the exclusion of $Z_i$ in the output for location $i$ (\textit{i.e.}, the so-called blind-spot network (BSN)). 

Given $m$ distinct (normalized, transformed) noisy images $\mathcal{Z}=\{\bm Z^{(j)}\}_{j=1}^m$, the training of BP-AIDE is done by minimizing 
\begin{equation}
    \mathcal{L}(\mathcal{Z};\wb) = \frac{1}{m}\sum_{j=1}^m \mathbf{L}_n\Big(\bm Z^{(j)},\bm f(\bm Z^{(j)};\wb);(\beta^{(j)})^{-2}\Big),\label{eq:BP_AIDE_loss}
\end{equation}
in which the $\mathbf{L}_n(\cdot;\cdot)$ in (\ref{eq:BP_AIDE_loss}) is defined in \cite{(NAIDE)cha2018neural} as
\begin{align}
\mathbf{L}_{n}(\bm Z,\bm f; {\sigma}^{2}) \triangleq  \frac{1}{n}{ \| \bm Z- \bm f\|_{2}^{2}} + \frac{{\sigma}^{2}}{n} \sum_{i=1}^{n}(2{a}_{1,i}-1),\label{eq:est_loss}
\end{align}
in which $\bm f\triangleq \bm f(\bm Z;\wb)$ and $a_{1,i}\triangleq a_{1}(\wb, \Cb_{k\times k}^{-i})$ for notational brevity. Now, it is shown in \cite[Lemma 1]{(FCAIDE)cha2019fully} that if the noise that generates $\bm Z$ is additive, independent, zero-mean with homogeneous variance, \textit{e.g.}, AWGN, then (\ref{eq:est_loss}) becomes an unbiased estimate of the mean-squared error (MSE) of $\bm f(\bm Z;\wb)$ for estimating the underlying clean image. An important point to emphasize is that such unbiasedness only holds for the pixelwise affine denoisers of the form (\ref{eq:affine}) with $\{a_m(\wb,\Cb_{k\times k}^{-i})\}_{m=0,1}$ being conditionally independent of $Z_i$ given $\bm Z^{-i}$ (\textit{i.e.}, the outputs of a BSN). Since the GAT transformed and normalized image $\bm Z^{(j)}$ approximately has AWGN with variance $(\beta^{(j)})^{-2}$, minimizing (\ref{eq:BP_AIDE_loss}), which only depends on the noisy images $\mathcal{Z}$ and no underlying cleans, indeed becomes minimizing the unbiased estimate of MSE. In \cite{(BPAIDE)9117146}, BP-AIDE simply adopted the BSN architecture proposed in \cite{(FCAIDE)cha2019fully} and trained the network for obtaining the pixelwise slope and intercept $\{a_m(\wb,\Cb_{k\times k}^{-i})\}_{m=0,1}$ for all $i$.

\noindent\textbf{(3) Inference} Once the training is done, when denoising a given test noisy image $\bm Y_{\text{te}}$ at the inference time, the Est+GAT step is first applied to obtain $\bm Z_{\text{te}}$, then gets denoised by $\bm f(\bm Z_{\text{te}},\wb_{\text{BP-ADIE}})$ in the transformed domain. Then, we apply the reverse of the normalization step in Est+GAT to obtain $\bm D$ and finally apply the IAT (\ref{eq:IAT}) with the estimated noise parameters $(\hat{\alpha},\hat{\sigma})$ obtained in the Est+GAT step.


\subsection{Motivation}
While BP-AIDE achieved the state-of-the-art denoising performance for real-world noise benchmarks among the methods that only use single noisy images, we observe its critical bottleneck in inference time lies in the Est+GAT step described above. Namely, the Poisson-Gaussian noise parameters $(\alpha,\sigma)$ need to be estimated for each and every given test image, which typically takes in the order of a few seconds using modern CPUs. Moreover, the BSN from \cite{(FCAIDE)cha2019fully}, denoted by QED network, uses three different kind of masked convolution filters, which results in large memory usage and unnecessary FLOPs that slows down the inference time. As described in the next section, we dramatically improve the inference time of BP-AIDE ($\times 10$) by learning to estimate the Poisson-Gaussian noise parameters with a separate network (PGE-Net) and devising a novel BSN (FBI-Net) with much simpler structure.



\section{Main Method} 
As illustrated in Figure \ref{fig:overall_process}, our proposed FBI-Denoiser consists of two phases: first train PGE-Net for Poisson-Gaussian noise estimation (\textit{Phase 1}), then train FBI-Net for blind denoising  (\textit{Phase 2}). In the subsequent sections, we first present the intuition of and the training procedure of PGE-Net (Section \ref{subsec:pge-net}), then present in details about the new BSN architecture (Section \ref{subsec:fbi-net}).

\subsection{Training PGE-Net:  \textbf{\textit{Phase 1 }} } 
\label{subsec:pge-net}
As mentioned in Section \ref{sec:gat}, GAT \cite{(GAT)anscombe1948transformation} is known to stabilize the noise such that the noise for each pixel in the transformed image, $\bm G_{\alpha,\sigma}(\bm Y)$, becomes approximately independent Gaussian with unit variance. We use this property of GAT to build an intuition for training a network to estimate the Poisson-Gaussian noise parameters, $(\alpha,\sigma)$. 

Before describing our method, we introduce a few more notations. First, let $\bm Y$ denote a noisy image as before that is corrupted by the Poisson-Gaussian noise model (\ref{eq:noise model}) with true parameters $(\alpha,\sigma)$. Moreover, define $B$ as a patch extracted from $\bm Y$ and $p_B(\bm x)$ as the PDF of the clean $\bm x$ over $B$. Moreover, denote $G_{\hat{\alpha},\hat{\sigma}}(\bm Y)$ as the GAT-transformed image with estimated $(\hat{\alpha},\hat{\sigma})$, and the stabilized noise variance of $\bm G_{\hat{\alpha},\hat{\sigma}}(\bm Y)$ is denoted by $\text{Var} \, (G_{\hat{\alpha},\hat{\sigma}}(\bm Y)|\bm x)$. Now, in \cite[Proposition 1]{makitalo2014noise}, it has been shown that, under reasonable assumptions, the following set
\begin{equation} 
\mathcal{S}_{B}\triangleq\Big\{(\hat{\alpha},\hat{\sigma}): \int \, \sqrt{\text{Var} \, (G_{\hat{\alpha},\hat{\sigma}}(\bm Y)\,| \,\bm x)} p_{B}(\bm x) dx =1 \Big\}\nonumber
\label{eq:VST_proposition}
\end{equation}
becomes a locally smooth curve around the true noise parameters $(\alpha,\sigma)$. With this proposition, \cite{makitalo2014noise} showed that when multiple patches, $\{B_i\}$'s, are extracted from an image, then the true $(\alpha,\sigma)$ typically lie in the intersection $\cap_{i}\mathcal{S}_{B_i}$.

\begin{figure}[h]
 \vspace{-.1in}
    \centering
    \subfigure{
    \includegraphics[width=0.9\columnwidth]{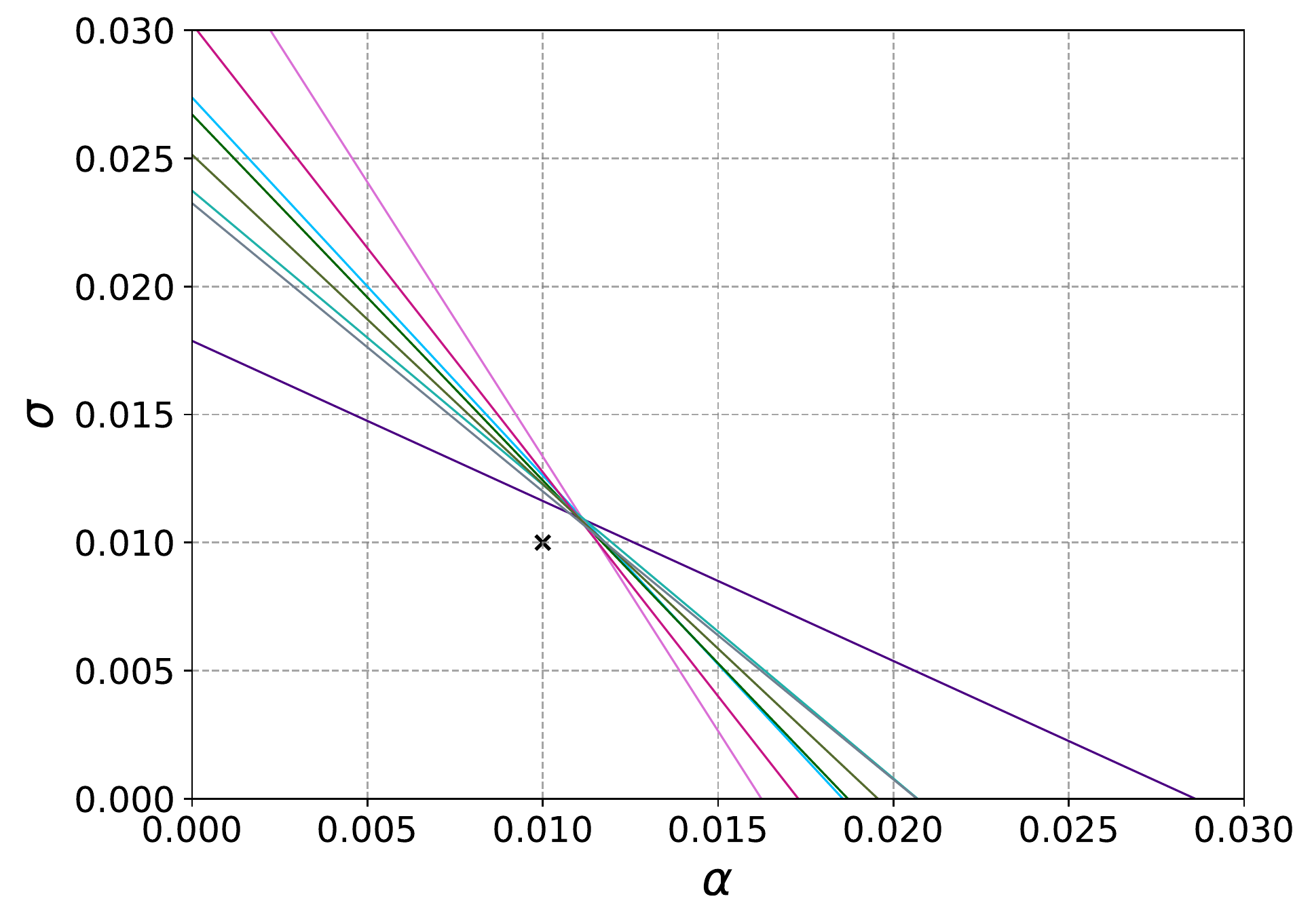}}
    \caption{$\mathcal{S}_B$ curves for ten patches each cropped from different images randomly selected from the MIT-Adobe FiveK Dataset \, \cite{(Fivek)bychkovsky2011learning} (True $(\alpha,\sigma)=(0.01,0.01)$ denoted by x.)}
    \label{fig:est_1_result}
 \vspace{-.2in}
\end{figure}

While the findings of \cite{makitalo2014noise} can be applied to estimating true $(\alpha,\sigma)$ for a single image, we extend this intuition for devising a neural network-based estimation method that can learn the estimation function from multiple noisy images. In Figure \ref{fig:est_1_result}, we show that when the patches $\{B_i\}$'s are extracted from multiple noisy images that are corrupted with Poisson-Gaussian noise with the same parameters $(\alpha,\sigma)$, the true $(\alpha,\sigma)$ also tends to lie near the intersection $\cap_i S_{B_i}$. This observation suggests that we can use patches from multiple noisy images and may learn a neural network model that can directly estimates the noise parameters $(\hat{\alpha},\hat{\sigma})$ that make the noise variance of $G_{\hat{\alpha},\hat{\sigma}}(\bm Y)$
close to $1$. Furthermore, when sufficiently many noisy images with various levels of $(\alpha,\sigma)$ are available as a training set, we may also expect the neural network to generalize well to unseen noise parameters.

Our PGE-Net exactly realizes the intuition made above. We first define a neural network, $h(\cdot,\bm\theta):\bm Y \rightarrow \mathbb{R}^2$, that takes the Poisson-Gaussian noise corrupted image $\bm Y$ and outputs the noise parameter estimates $(\hat{\alpha},\hat{\sigma})$. Moreover, we denote $\eta(\cdot):\bm Z\rightarrow \mathbb{R}$ as the function in \cite{(chen)chen2015efficient} that estimates the Gaussian noise variance from an input image $\bm Z$. Then, given $m$ distinct noisy images $\mathcal{Y}=\{\bm Y^{(j)}\}_{j=1}^m$, the loss function for our PGE-Net
becomes 
\begin{align}
\mathcal{L}_{\text{PGE}}(\mathcal{Y};\bm\theta) \triangleq \sum_{j=1}^m \Big(\mathit{\eta} \big(G_{\hat{\alpha}(\bm\theta), \hat{\sigma}(\bm\theta)}(\bm Y^{(j)})\big )- 1\Big)^2,\label{eq:loss_noise_est}
\end{align}
in which $\hat{\alpha}(\bm\theta) \triangleq h_1(\bm Y^{(j)}; \bm\theta)$ and $\hat{\sigma}(\bm\theta) \triangleq h_2(\bm Y^{(j)}; \bm\theta)$ are the estimated noise parameters $(\hat{\alpha},\hat{\sigma})$ that are outputs of $h(\cdot,\bm\theta)$. For the specific architecture of $h(\cdot,\bm\theta)$, we used a three layer U-Net \cite{(UNet)ronneberger2015u} architecture and average pooling layer in the output layer. 

Now, when using (\ref{eq:loss_noise_est}) as a loss function to minimize for training $h(\cdot,\bm\theta)$, an important point to check is whether all the operations for computing the loss is differentiable with respect to the network parameter $\bm\theta$ and can be implemented with tensor operations. To that end, we can see that the GAT (\ref{eq:gat}) can be tensorized by a linear operation and applying an element-wise function. However, implementing the Gaussian noise variance estimation $\eta(\cdot)$ \cite{(chen)chen2015efficient} with tensor operations is not trivial since it involves iterative for-loop procedures. To that end, we utilized several tricks using the lower triangular matrix multiplication and masking scheme to replace iterative procedures with tensor operations. Details about the implementations can be found in the Supplementary Material (S.M).

\textcolor{\tsmcolor}{
Note the main purpose of PGE-Net is not to simply estimate the noise parameters in a myopic way, but to estimate them \textit{fast} such that the GAT-transformed image (with the estimated parameters) has stabilized noise variance close to 1 for the next denoisng step.} Once our PGE-Net is learned via minimizing (\ref{eq:loss_noise_est}), it enjoys extremely fast inference time (using GPU) as only simple forward pass of the neural network is required to estimate the noise parameters $(\hat{\alpha},\hat{\sigma})$. Moreover, as we show in Section \ref{subsec: experiment_noise_estimation}, our method becomes much more stable than the conventional noise estimation methods since PGE-Net can smoothly generalize from multiple noisy images, whereas the methods in \cite{(Foi)foi2008practical,(pg_Liu)liu2014practical}, which run optimization routines separately for each image,  may fail to correctly estimate depending on the input image. 

\begin{figure*}[t]
    \centering
     \subfigure[HQ-SSL \cite{(HQDenoising)laine2019high}]{\label{fig:hq}
    \includegraphics[width=0.19\linewidth]{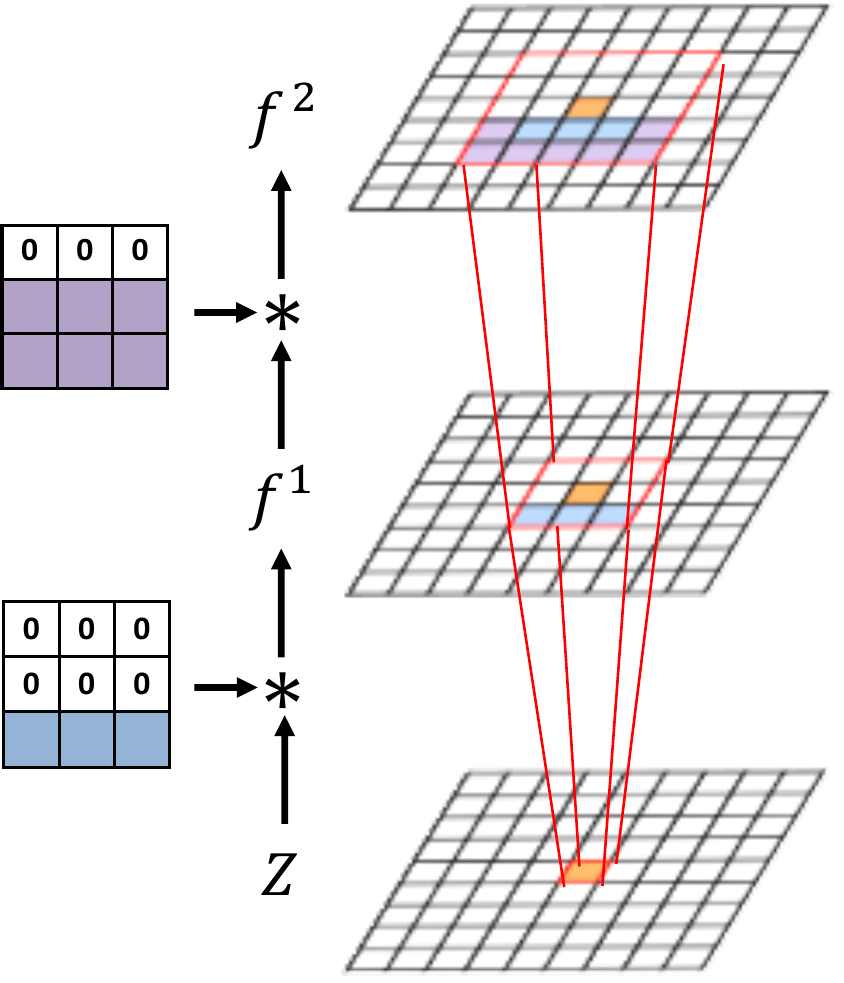}}
     \subfigure[D-BSN \cite{(Unpaired)wu2020unpaired}]{\label{fig:dbsn}
    \includegraphics[width=0.22\linewidth]{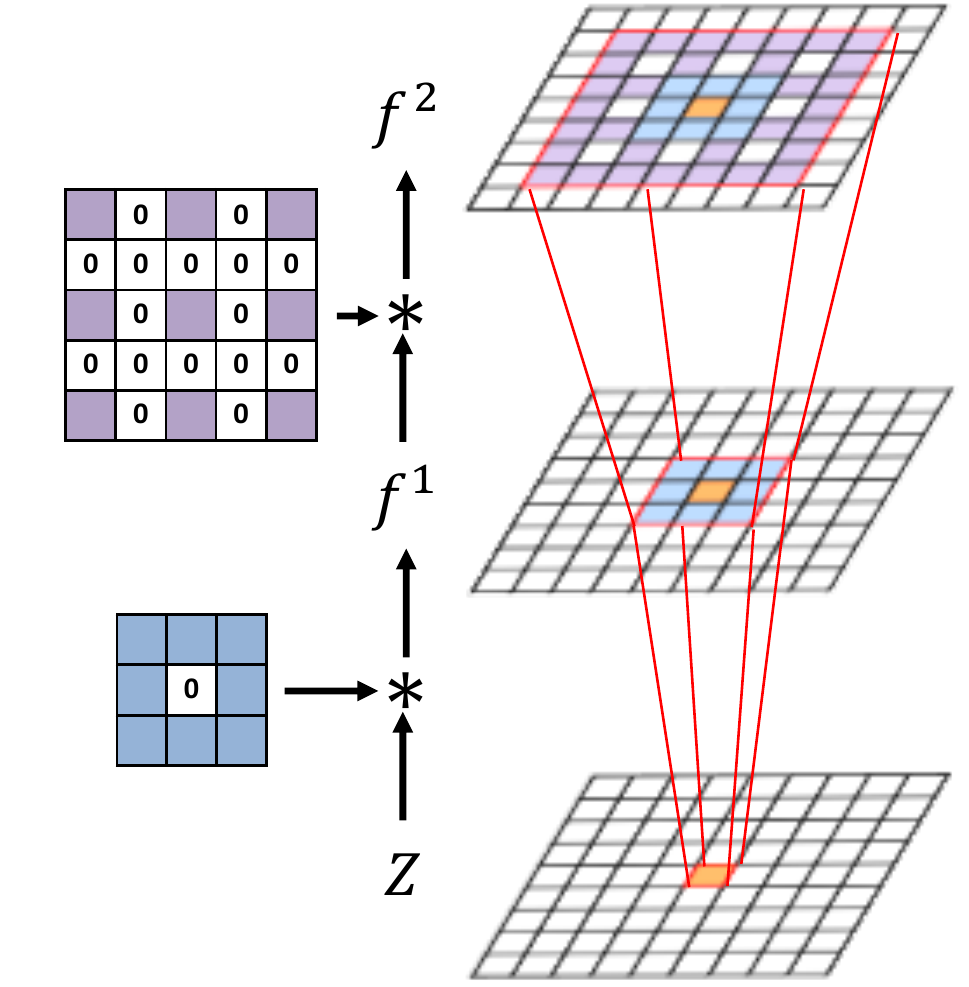}}
     \subfigure[FC-AIDE \cite{(FCAIDE)cha2019fully}]{\label{fig:fcaide}
    \includegraphics[width=0.225\linewidth]{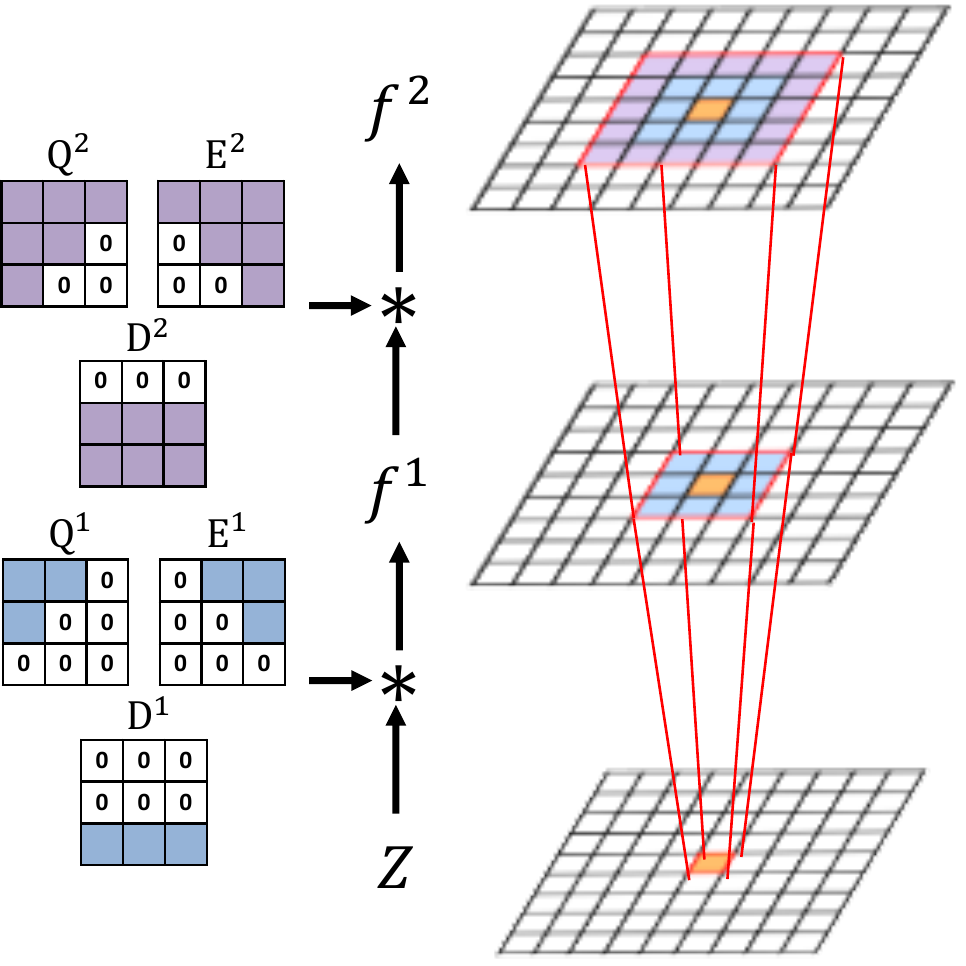}}
     \subfigure[FBI-Net]{\label{fig:fbi_layers}
    \includegraphics[width=0.25\linewidth]{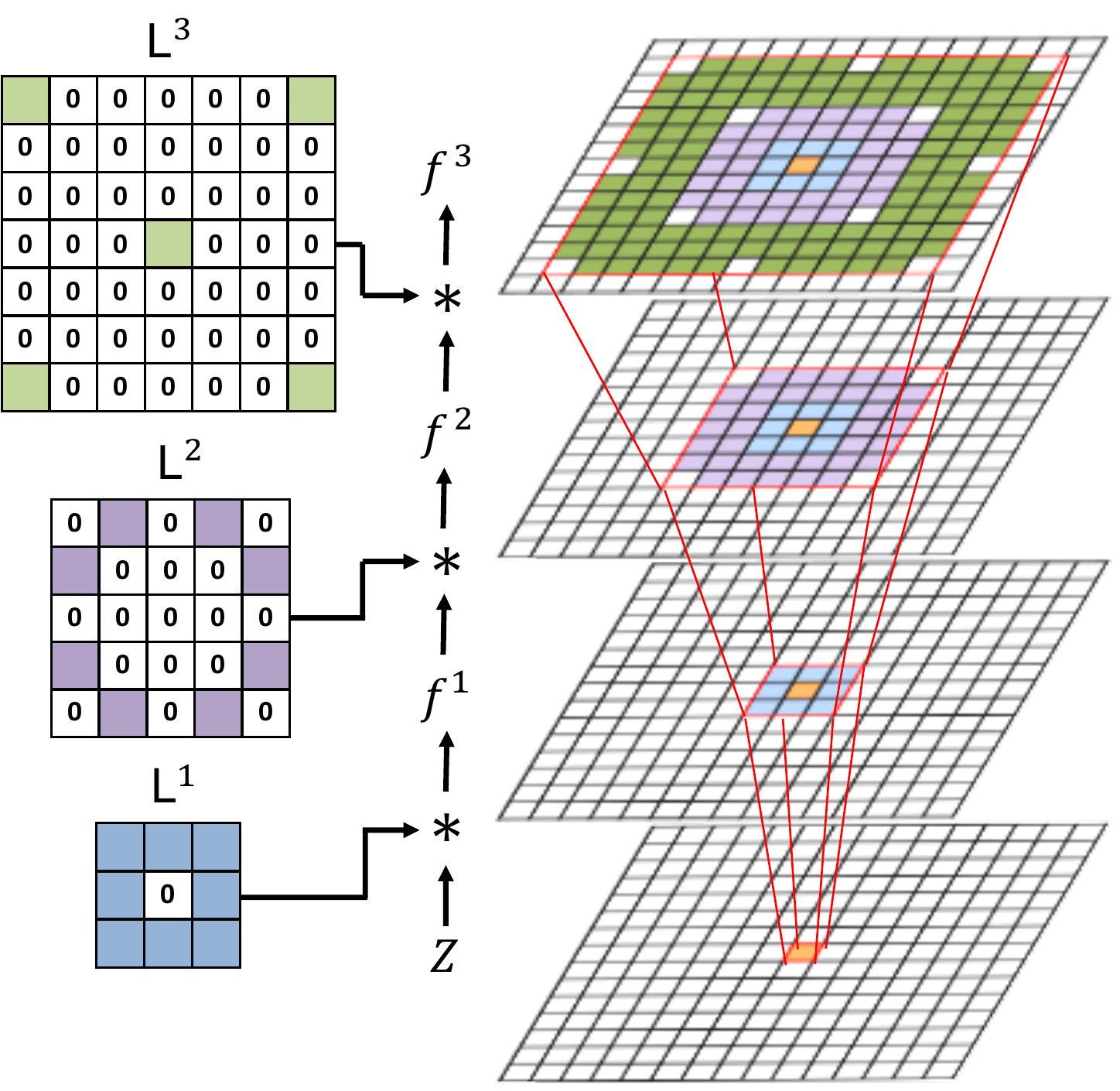}}
        \subfigure[Overall architecture of FBI-Net]{\label{fig:overall}
    \includegraphics[width=0.78\linewidth]{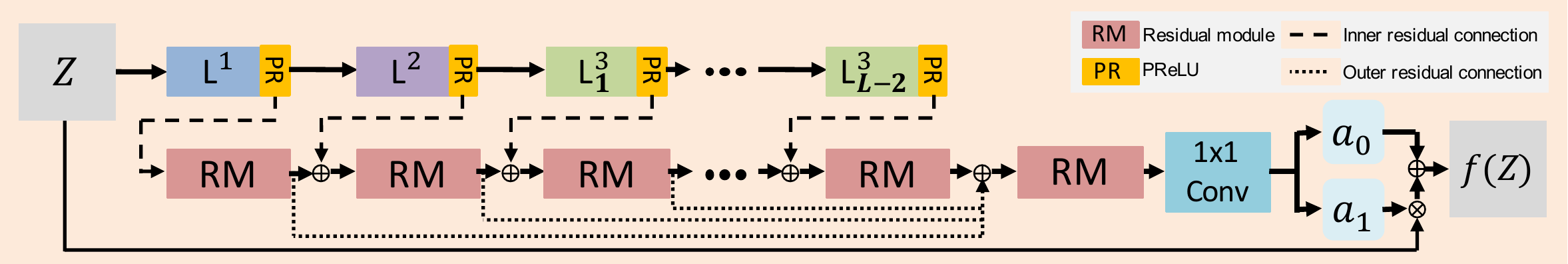}}
        \subfigure[Residual Module]{\label{fig:residual}
    \includegraphics[width=0.2\linewidth]{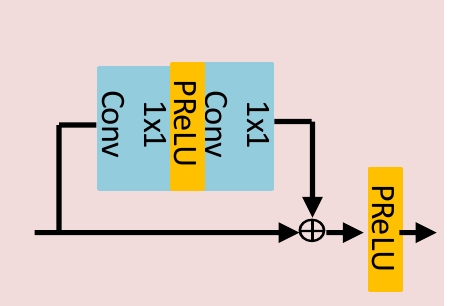}}
    \caption{The layers of previously proposed blind spot network and our FBI-Net.}
    \label{fig:overall_architecture}\vspace{-.1in}
\end{figure*}
\subsection{Training FBI-Net: \textbf{\textit{Phase 2}} }\label{subsec:fbi-net}
Once the noise parameters $(\hat{\alpha},\hat{\sigma})$ are estimated with PGE-Net, our denoising network FBI-Net is trained following the same training procedure of BP-AIDE described in Section \ref{subsec:bp-aide} (as also shown in \textit{Phase 2} of Figure \ref{fig:overall_process}); namely, minimize (\ref{eq:BP_AIDE_loss}) with the transformed (with the estimated $(\hat{\alpha},\hat{\sigma})$) and normalized images $\bm Z^{(j)}$. Recall that the BSN architecture was necessary for implementing the pixelwise affine denoiser (\ref{eq:affine}) and maintaining the unbiasedness of (\ref{eq:est_loss}), and we devise a more efficient BSN in this section. 

The ideal condition for BSN is to have a simple structure that excludes only $Z_i$ when computing the activation value at location $i$ in any layer. In practice, some methods implemented BSN by discarding some more pixels other than $Z_i$ when computing the activation value at location $i$, but obviously, the more we exclude, the worse the performance becomes. 
Recently, several network architecture to enhance the efficiency of BSN were proposed
\cite{(FCAIDE)cha2019fully, (N2V)krull2018noise2void, (HQDenoising)laine2019high, (Unpaired)wu2020unpaired} in the denoising context.  
N2V \cite{(N2V)krull2018noise2void} maintains above constraint indirectly by adopting randomly masking scheme. The masking scheme severely degrades training efficiency since only few pixels can be utilized. 
HQ-SSL \cite{(HQDenoising)laine2019high} suggests a more efficient constrained model by adopting four-way rotations as shown in Figure \ref{fig:hq}. But the requirement of four rotated input images gives limited efficiency.
D-BSN \cite{(Unpaired)wu2020unpaired} devised a completely re-designed model. As shown in Figure \ref{fig:dbsn}, they mainly use two different types of a masked CNN layer and $1 \times1$ convolution layer to maintain the constraint. This model significantly improves efficiency over previous methods, however, it still has the disadvantage of having a lot of unwanted holes excluded from the receptive field.
As shown in Figure \ref{fig:fcaide}, FC-AIDE \cite{(FCAIDE)cha2019fully} designed three different layers, denoted as Q, E and D layer. By utilizing three layers, they obtain the ideal condition we described above. However, since each of the three layers has to be stacked separately for getting it, the model is relatively complex and its inference time is slow.

 We propose a more efficient BSN architecture (FBI-Net) by re-designing three different zero-masked convolution layers, L$^1$, L$^2$ and L$^3$ as described in Figure \ref{fig:fbi_layers}. L$^1$ is a center masked $3\times3$ size of convolution layer which is equal with the first layer proposed in D-BSN. L$^2$ is $5\times5$ size of convolution layer which is all masked except for eight holes and L$^3$ is $7\times7$ size of convolution layer which only has weights at the center and each edge. Note that L$^3$ can be replaced with $3\times3$ size of a dilated convolution layer (dilation = 3). By stacking these layer sequentially, we can get an almost full receptive field with fewer holes than D-BSN. Compared with FC-AIDE, even though FBI-Net have some holes in the receptive field, our layers have the advantage of obtaining a simpler architecture and fewer parameters.
 \begin{table}[h!]
\vspace{-.1in}
\caption{The comparison of BSN on an image of size 512$\times$512.}
\centering
\smallskip\noindent
\resizebox{.7\linewidth}{!}{
\begin{tabular}{|c||c|c|c|}
\hline
                                                                 & FC-AIDE & D-BSN     & \textbf{FBI-Net} \\ \hline \hline
Num of parameters                                                & 754,000 & 6,612,000 & \textbf{340,000} \\ \hline
\begin{tabular}[c]{@{}c@{}}GPU memory\\ requirement\end{tabular} & 2,581MB & 4,231MB   & \textbf{2,512MB} \\ \hline
Inference time                                                   & 0.29    & 0.99      & \textbf{0.21}    \\ \hline
\end{tabular}}
  \vspace{-.1in}
\label{table:ablation_blind_spot}
\end{table}
Figure \ref{fig:overall_architecture} shows the overall architecture of FBI-Net. We sequentially stacked layers in the order of L$^1$, L$^2$ and L$^3$ but we only stacked L$^3$ after $4$-th layer. Following the finding of FC-AIDE \cite{(FCAIDE)cha2019fully}, we used PReLU \cite{(PReLU)he2015delving} for all activation functions and Residual Module (RM) proposed in Figure \ref{fig:residual}. To maintain the constraint, we only used $1\times1$ CNN layer for the output layer and RM. For all experiments, we set $L=17$ which has a receptive field of 181$\times$181 size. Note that we add two different residual connections denoted by Inner and Outer residual connection. We found out that these residual connections make more stable training and improve the denoising performance as can be shown in the ablation study. Table \ref{table:ablation_blind_spot} shows the performance comparison of each BSN when denoising an image of size $512\times512$. 
Compared to the baselines, the proposed FBI-Net uses much smaller number of parameters and achieves the fastest inference time while the GPU memory requirement is small. 

\section{Experimental Results}
\subsection{Data and Experimental settings}
\noindent\textbf{Data and Implementation details} \ \
In synthetic noise experiments, we used two datasets: BSD400 \cite{(BSD400)martin2001database} for grayscale images and MIT-Adobe FiveK Dataset (Fivek) \cite{(Fivek)bychkovsky2011learning} for raw-RGB images. 
For evaluation of the grayscale images, the standard BSD68 dataset \cite{(BSD68)roth2005fields} was used. 
For Fivek, which is composed of $5,000$ images, $4,800$ images were used for training, and the remaining $200$ were used as a test set. 
To reflect the real noise characteristics, various levels of Poisson-Gaussian noise were simulated. 
Note each model is individually trained and tested with designated noise characteristics.  
For real noise experiments, we used three real-world noisy image datasets: Fluorescence Microscopy Denoising (FMD) dataset \cite{(FMD)zhang2019poisson}, which is composed of grayscale microscopy images, and SIDD \cite{(SIDD)SIDD_2018_CVPR} / DND \cite{(DND)plotz2017benchmarking} which consist of raw-RGB and sRGB images.
In FMD, raw noisy images from three separate general configurations are used; confocal FISH (CF FISH), confocal MICE (CF MICE) and two-photon MICE (TP MICE). 
The training on SIDD and DND was done using only raw-RGB images, and we received the evaluation results on both raw-RGB and sRGB test sets by submitting the results to the public websites \cite{siddweb, dndweb}, respectively. 
Note that SIDD provides training, validation and test datasets, but DND only provides test images to the public.
For training a model, we generated $m= 25,000$ noisy patches of size $210 \times 210$ from each dataset. 
To reflect the complexity of real noise, we reduced the range of slope coefficient $a_{1}(\cdot)$ in (\ref{eq:affine}), \textit{i.e.}, we changed from $[0,1]$ which is the original range of $a_{1}(\cdot)$ suggested by \cite{(BPAIDE)9117146} to $[0,0.1]$. 
For a fair comparison, we used the same range $[0,0.1]$ for both BP-AIDE and FBI-Denoiser in all experiments. 
The details on the software platform, training and hyperparameters are in the S.M.


\noindent\textbf{Baselines} \ \
We first compared the accuracy of the noise parameters from PGE-Net (\textit{phase 1}) with two representative baselines: Foi \cite{(Foi)foi2008practical} and Liu \cite{(pg_Liu)liu2014practical}.
Then, we compared the overall denoising performance of FBI-Denoiser with following baselines: GAT+BM3D \cite{(BM3D)dabov2007image}, N2V \cite{(N2V)krull2018noise2void}, D-BSN \cite{(Unpaired)wu2020unpaired}, and BP-AIDE \cite{(BPAIDE)9117146}. 
GAT+BM3D is a traditional method, but it is still a \textit{very} powerful baseline for the Poisson-Gaussian noise denoising.
The estimation method in Liu \cite{(pg_Liu)liu2014practical} is used as a noise estimation method for GAT+BM3D and BP-AIDE, since it achieves robust performance in real-world noise benchmarks. 
Moreover, the self-supervised step of D-BSN is used as a baseline for fair comparison.
As an upper bound, we trained a model in a supervised way (\textit{i.e.}, using clean target images), denoted as "Sup", that uses the same network architecture as FBI-Net. 
All results of the baselines are reproduced with publicly available source codes, and for brevity, FBI-Denoiser is denoted as ``FBI-D'' in this section.

\subsection{Experimental results on noise estimation}\label{subsec: experiment_noise_estimation}
Here, we first validate the effectiveness of the noise estimation of PGE-Net. Table \ref{table:result_est_mean_std} shows the \textit{average} of the estimated $(\hat{\alpha}, \hat{\sigma})$ values, obtained by Foi \cite{(Foi)foi2008practical}, Liu \cite{ (pg_Liu)liu2014practical} and PGE-Net from BSD68 images that are corrupted by four different noise levels specified by each row in the table. 
\begin{table}[h]
\caption{The average values of estimated $\hat{\alpha}$ and $\hat{\sigma}$ for BSD68. \textbf{Bold} denotes the best result among the three methods.}
\centering
\smallskip\noindent
\resizebox{.7\linewidth}{!}{

\begin{tabular}{|c||c|c|c|c|c|c|}
\hline
\multirow{2}{*}{ \begin{tabular}[c]{@{}c@{}} Noise level \\ $(\alpha, \sigma) $\end{tabular}} & \multicolumn{2}{c|}{Foi \cite{(Foi)foi2008practical}} & \multicolumn{2}{c|}{Liu \cite{(pg_Liu)liu2014practical}} & \multicolumn{2}{c|}{PGE-Net} \\ \cline{2-7} 
                             &$\hat{\alpha}$             &$\hat{\sigma}$            &$\hat{\alpha}$             &$\hat{\sigma}$          &$\hat{\alpha}$             &$\hat{\sigma}$           \\ \hline \hline
(0.1, 0.02)                            & \begin{tabular}[c]{@{}c@{}}$0.096$\end{tabular}               &\begin{tabular}[c]{@{}c@{}}$\textbf{0.042}$\end{tabular}                &\begin{tabular}[c]{@{}c@{}}$0.072$\end{tabular}              & \begin{tabular}[c]{@{}c@{}}$0.045$\end{tabular}   & \begin{tabular}[c]{@{}c@{}}$\textbf{0.098}$\end{tabular}              & \begin{tabular}[c]{@{}c@{}}$0.003$\end{tabular}           \\ \hline

(0.1, 0.0002)                   & \begin{tabular}[c]{@{}c@{}}$\textbf{0.097}$\end{tabular}               & \begin{tabular}[c]{@{}c@{}}$0.035$\end{tabular}   & \begin{tabular}[c]{@{}c@{}}$0.071$ \end{tabular}             &\begin{tabular}[c]{@{}c@{}}$0.044$\end{tabular}            & \begin{tabular}[c]{@{}c@{}}$0.095$\end{tabular}              & \begin{tabular}[c]{@{}c@{}}$\textbf{0.0001}$\end{tabular}        \\ \hline

(0.05, 0.02)                   & \begin{tabular}[c]{@{}c@{}}$\textbf{0.049}$\end{tabular}               & \begin{tabular}[c]{@{}c@{}}$\textbf{0.031}$\end{tabular}   & \begin{tabular}[c]{@{}c@{}}$0.04$ \end{tabular}             &\begin{tabular}[c]{@{}c@{}}$0.04$\end{tabular}            & \begin{tabular}[c]{@{}c@{}}$0.052$\end{tabular}              & \begin{tabular}[c]{@{}c@{}}$0.0001$\end{tabular}        \\ \hline

(0.05, 0.0002)                   & \begin{tabular}[c]{@{}c@{}}$\textbf{0.051}$\end{tabular}               & \begin{tabular}[c]{@{}c@{}}$0.018$\end{tabular}   & \begin{tabular}[c]{@{}c@{}}$0.039$ \end{tabular}             &\begin{tabular}[c]{@{}c@{}}$0.034$\end{tabular}            & \begin{tabular}[c]{@{}c@{}}$0.051$\end{tabular}              & \begin{tabular}[c]{@{}c@{}}$\textbf{0.0001}$\end{tabular}        \\ \hline

\end{tabular}
}
\label{table:result_est_mean_std}
\end{table}
In addition, we indirectly measured the performance of noise estimation by comparing the denoising performance (on BSD68 and Fivek) of GAT+BM3D that uses the estimated noise parameters from each estimation method, as shown in Table \ref{table:result_estimation_GAT+BM3D}. Moreover, the performance of GAT+BM3D using the ground truth noise level is reported in the table as an upper bound. 
\textcolor{\tsmcolor}{Firstly, from Table \ref{table:result_est_mean_std}, we observed that unlike for $\alpha$, PGE-Net seems to significantly underestimate $\sigma$. 
However, we observe from Table \ref{table:result_estimation_GAT+BM3D} that  GAT+BM3D with the estimated parameters of PGE-Net 
still shows competitive denoising performance compared to others \cite{(Foi)foi2008practical, (pg_Liu)liu2014practical} with \textit{much faster} inference time ($\times 2000$). 
This suggests that the underestimated $\hat{\sigma}$ of PGE-Net has little impact on the denoising performance of GAT+BM3D or BP-AIDE framework (which carry out GAT using $\hat{\sigma}$).
} 
 \begin{figure}[h]
     \centering
     \subfigure[$(\alpha = 0.1 \,, \sigma=0.02)$ ]{
     \includegraphics[width=0.48\columnwidth]{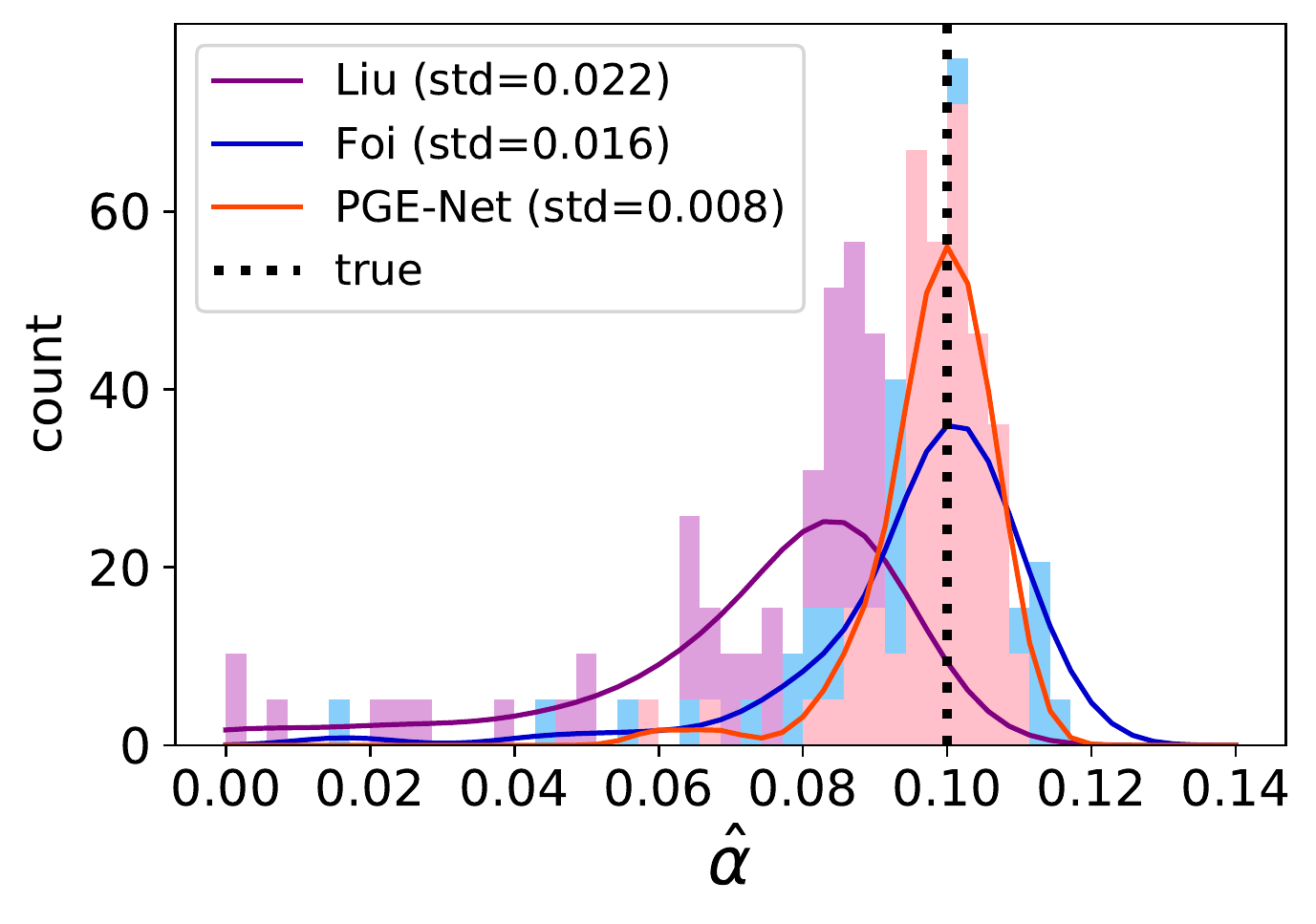}}
    \subfigure[$(\alpha = 0.05\,, \sigma=0.0002)$]{
     \includegraphics[width=0.48\columnwidth]{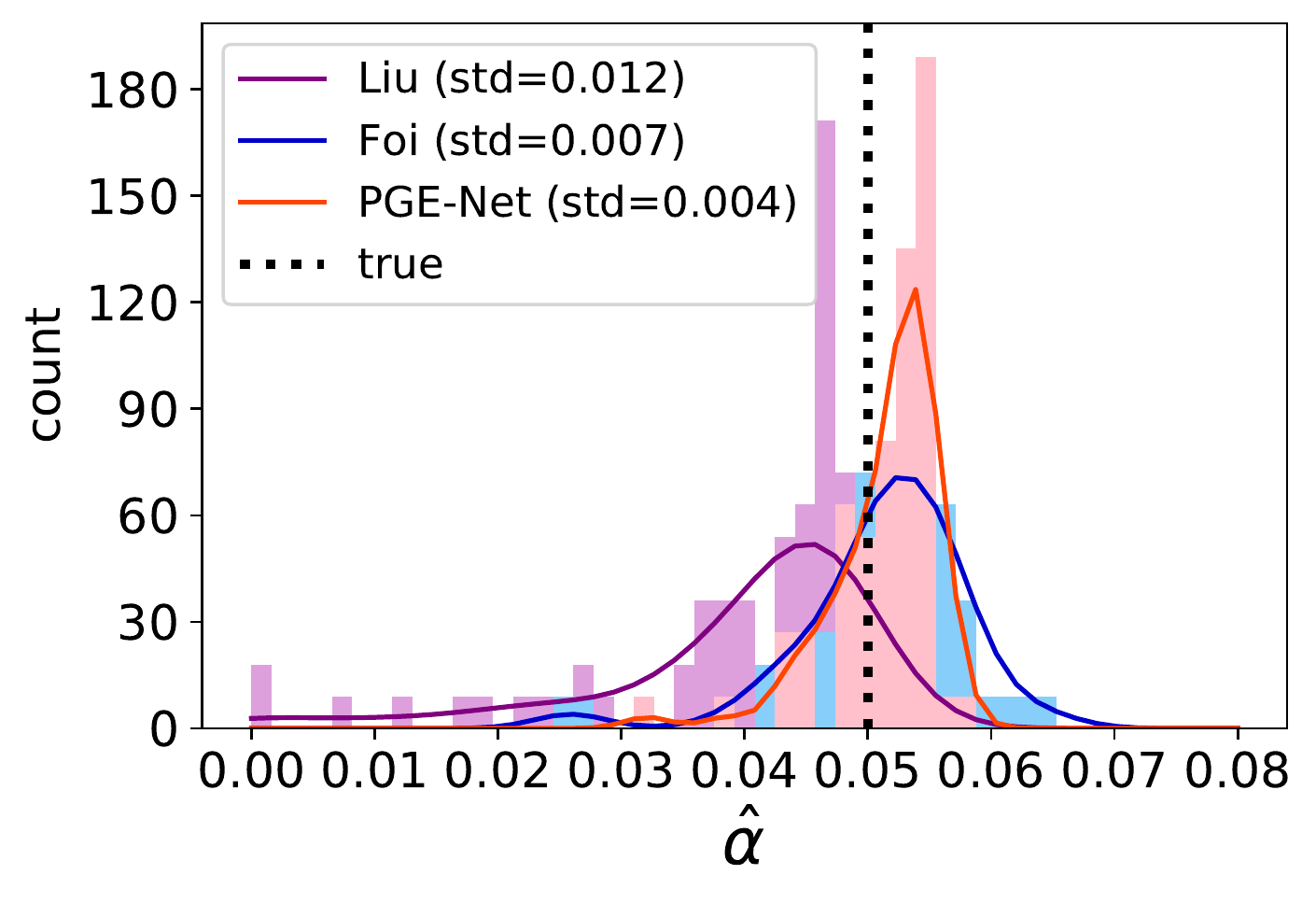}}
         \vspace{-.1in}
     \caption{Histogram of estimated $\hat{\alpha}$ for BSD68 from each estimation method. }\label{fig:mean_std_est_result}
     \label{fig:stable_result}
 \end{figure}   
\textcolor{\tsmcolor}{As we emphasized in Section \ref{subsec:pge-net}, we believe this somewhat counter-intuitive result is possible since
the underestimation of PGE-Net for $\sigma$ turns out to be \textit{inconsequential} for ensuring that the GAT-transformed image (with the estimated parameters of PGE-Net) has stabilized noise with homogeneous variance close to 1. 
To verify this more in detail, we conducted an experiment using an additional toy example in S.M.}
\begin{table}[h!]
 \vspace{-.1in}
\caption{Denoising results of GAT+BM3D with the estimated noise parameters. Red and blue denote the highest and the second highest result among the noise estimation algorithms, respectively.
}
\centering
\smallskip\noindent
\resizebox{\linewidth}{!}{
\begin{tabular}{|c|c|c||c|c|c||c|}
\hline
\multicolumn{3}{|c||}{\multirow{2}{*}{\begin{tabular}[c]{@{}c@{}}Dataset \\ (PSNR / SSIM)\end{tabular}}}                                 & \multicolumn{4}{c|}{\begin{tabular}[c]{@{}c@{}}Performance of GAT+BM3D\end{tabular}} \\ \cline{4-7} 
\multicolumn{3}{|c||}{}                                                         & Foi                             & Liu                             & PGE-Net           & Ground truth            \\ \hline \hline
\multirow{5}{*}{BSD68} & \multirow{4}{*}{$(\alpha, \sigma)$} & (0.01,0.0002)   & 29.88 / 0.8432                  & \textbf{\color{red}{30.33}} / \textbf{\color{red}{0.8623}}
                 & \textbf{\color{blue}{30.14}} / \textbf{\color{blue}{0.8521}}          & 30.34 / 0.8641      \\ \cline{3-7} 
                       &                                     & (0.01,0.02)     & 29.73 / 0.8393                  & \textbf{\color{red}{30.08}} / \textbf{\color{red}{0.8564}}               & \textbf{\color{blue}{29.81}} / \textbf{\color{blue}{0.8433}}       & 30.16 / 0.8570         \\ \cline{3-7} 
                       &                                     & (0.05,0.0002)   & 26.07 / 0.7292                  & \textbf{\color{red}{26.16}} / \textbf{\color{red}{0.7371}}                  & \textbf{\color{blue}{26.14}} / \textbf{\color{blue}{0.7335}}        & 26.18 / 0.7362        \\ \cline{3-7} 
                       &                                     & (0.05,0.02)     & 26.03 / 0.7269                  & \textbf{\color{red}{26.12}} / \textbf{\color{red}{0.7352}}                  & \textbf{\color{blue}{26.11}} / \textbf{\color{blue}{0.7344}}         & 26.16 / 0.7349       \\ \cline{2-7} 
                       & \multicolumn{2}{c||}{Time for noise estimation}                   & 3.123s                          & \textbf{\color{blue}{1.084s}}                          & \textbf{\color{red}{0.002s}}           & -             \\ \hline \hline
\multirow{5}{*}{Fivek} & \multirow{4}{*}{$(\alpha, \sigma)$} & (0.0005,0.0002) & 43.63 / 0.8645                  & \textbf{\color{blue}{49.05}} / \textbf{\color{blue}{0.9722}}                  & \textbf{\color{red}{49.11}} / \textbf{\color{red}{0.9736}}     & 49.41 / 0.9764            \\ \cline{3-7} 
                       &                                     & (0.0005,0.02)   & 41.90 / 0.8586                  & \textbf{\color{red}{44.51}} / \textbf{\color{red}{0.9188}}                  & \textbf{\color{blue}{44.11}} / \textbf{\color{blue}{0.9144}}       & 44.57 / 0.9173               \\ \cline{3-7} 
                       &                                     & (0.01,0.0002)   & 40.24 / 0.8493                  & \textbf{\color{red}{42.33}} / \textbf{\color{red}{0.9170}}                  & \textbf{\color{blue}{42.01}} / \textbf{\color{blue}{0.9060}}        & 42.46 / 0.9152  \\ \cline{3-7} 
                       &                                     & (0.01,0.02)     & 40.95 / 0.8347                  & \textbf{\color{blue}{41.26}} / \textbf{\color{blue}{0.8857}}                  & \textbf{\color{red}{41.28}} / \textbf{\color{red}{0.8863}}           & 41.14 / 0.8812      \\ \cline{2-7} 
                       & \multicolumn{2}{c||}{Time for noise estimation}                   & 2.521s                          & \textbf{\color{blue}{1.812s}}                          & \textbf{\color{red}{0.002s}}                       & - \\ \hline
\end{tabular}    
}
\label{table:result_estimation_GAT+BM3D}
  \vspace{-.15in}
\end{table}

%
%
Furthermore, Figure \ref{fig:mean_std_est_result} shows the histograms of the $\hat{\alpha}$ values (on BSD68) for two different noise levels obtained by each estimation method. From the figure, we observe that the standard deviations of $\hat{\alpha}$'s obtained by PGE-Net are far smaller than those of others, and, in particular, the failure case (in which $\hat{\alpha}$ goes to zero) does not occur in PGE-Net. 
Note such failure case of $\hat{\alpha}$ is fatal for training BP-AIDE or FBI-Denoiser since the extremely small value of $\hat{\alpha}$, which is a denominator of GAT (\ref{eq:gat}), causes the gradient explosion.

\begin{figure*}[t]
    \centering
    \subfigure{
    \includegraphics[width=0.95\linewidth]{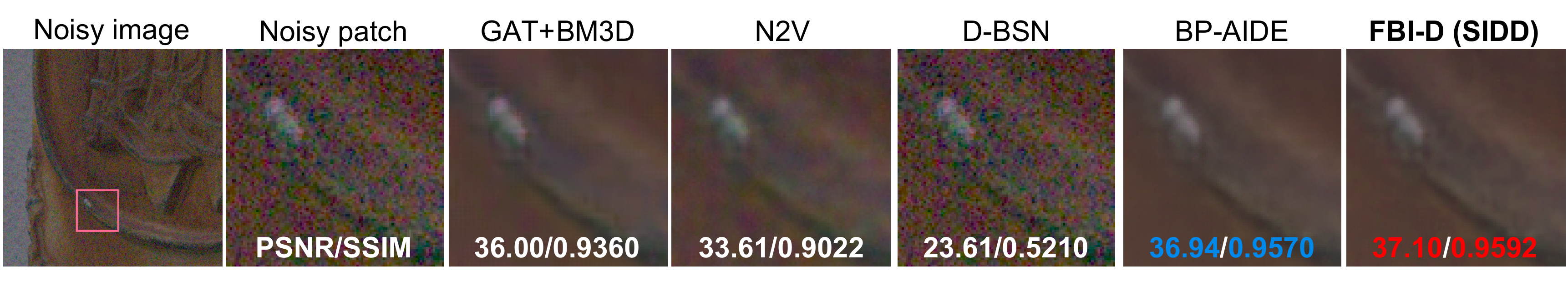}}
    \vspace{-.1in}
    \caption{Visualization results of DND}
    \label{fig:visualization}
    \vspace{-.2in}
\end{figure*}
\subsection{Experimental results on denoising}
\noindent\textbf{Synthetic noise} \ \ 
The experimental results on synthetic raw-RGB Fivek dataset are shown in the top of Table \ref{table:result_synthetic}.
\textcolor{\tsmcolor}{We simulated not only the specified levels of ($\alpha$, $\sigma$), but also a mixture of noise levels denoted as ``Mixture noise'', which were generated by following the same procedure as \cite{(UPI)brooks2019unprocessing} for sampling multiple noise levels.}
From the results, we first note that our FBI-D performs very well (often the best) compared to other baselines across various noise levels, including the ``Mixture noise'' case, with a fast inference time. 
From this result, we can confirm again that PGE-Net is very effective for FBI-D to achieve a competitive denoising performance.
Secondly, BP-AIDE and GAT+BM3D also aschieve good results, however, these methods have \textit{very} slow inference times due to the computational cost of the noise estimation method in Liu \cite{(pg_Liu)liu2014practical}.
Finally, D-BSN achieves a good performance for the weak noise cases, which surpasses FBI-D, however, when the noise is mixed or strong, D-BSN is far inferior to others.
In S.M, we analyze why FBI-D achieves relatively low performance in the weak noise cases. 
\begin{table}[h!]
\vspace{-.05in}
\caption{PSNR(dB)/SSIM on  Fivek (Synthetic) / FMD (Real) validation dataset. The colored texts are as before.}
\centering
\smallskip\noindent
\resizebox{.98\linewidth}{!}{
\begin{tabular}{|c|c|c||c|c|c|c|c||c|}
\hline
\multicolumn{3}{|c||}{Dataset}                                                                          & GAT+BM3D                                                & N2V                                                     & D-BSN                                                   & BP-AIDE                                                 & FBI-D                                                     & FBI-D (Sup)                                               \\ \hline \hline
\multirow{4}[10]{*}{\begin{tabular}[c]{@{}c@{}}Synthetic\end{tabular}}
&\multirow{4}[10]{*}{\begin{tabular}[c]{@{}c@{}}Fivek\\ ($\alpha$, $\sigma$)\end{tabular}} & (0.01, 0.0002) & \begin{tabular}[c]{@{}c@{}}42.33\\ /0.9170\end{tabular} & \begin{tabular}[c]{@{}c@{}}30.39\\ /0.8541\end{tabular} & \begin{tabular}[c]{@{}c@{}}\textbf{\color{red}{44.50}}\\ /\textbf{\color{red}{0.9602}}\end{tabular} & \begin{tabular}[c]{@{}c@{}}43.54\\ /0.9464\end{tabular} & \begin{tabular}[c]{@{}c@{}}\textbf{\color{blue}{44.43}}\\ /\textbf{\color{blue}{0.9569}}\end{tabular} & \begin{tabular}[c]{@{}c@{}}44.80\\ /0.9600\end{tabular} \\ \cline{3-9} 
                                                                         &             & (0.01, 0.02)   & \begin{tabular}[c]{@{}c@{}}41.26\\ /0.8857\end{tabular} & \begin{tabular}[c]{@{}c@{}}29.21\\ /0.8168\end{tabular} & \begin{tabular}[c]{@{}c@{}}38.06\\ /0.8280\end{tabular} & \begin{tabular}[c]{@{}c@{}}\textbf{\color{blue}{42.59}}\\ /\textbf{\color{blue}{0.9350}}\end{tabular} & \begin{tabular}[c]{@{}c@{}}\textbf{\color{red}{43.14}}\\ /\textbf{\color{red}{0.9402}}\end{tabular} & \begin{tabular}[c]{@{}c@{}}44.38\\ /0.9537\end{tabular} \\ \cline{3-9} 
                                                                         &             & (0.05, 0.02)   & \begin{tabular}[c]{@{}c@{}}36.42\\ /0.8034\end{tabular} &    \begin{tabular}[c]{@{}c@{}}25.84\\ /0.7386\end{tabular}                                                     & \begin{tabular}[c]{@{}c@{}}26.13\\ /0.4363\end{tabular} & \begin{tabular}[c]{@{}c@{}}\textbf{\color{blue}{38.37}}\\ /\textbf{\color{blue}{0.9107}}\end{tabular} & \begin{tabular}[c]{@{}c@{}}\textbf{\color{red}{39.30}}\\ /\textbf{\color{red}{0.9165}}\end{tabular} & \begin{tabular}[c]{@{}c@{}}42.17\\ /0.9402\end{tabular} \\ \cline{3-9} 
                                                                        &              & Mixture noise   & \begin{tabular}[c]{@{}c@{}}\textbf{\color{red}{46.94}}\\ /0.9560\end{tabular} &      \begin{tabular}[c]{@{}c@{}}33.43\\ /0.8949\end{tabular}                                                   & \begin{tabular}[c]{@{}c@{}}39.91\\ /0.8336\end{tabular} & \begin{tabular}[c]{@{}c@{}}46.43\\ /\textbf{\color{blue}{0.9658}}\end{tabular} & \begin{tabular}[c]{@{}c@{}}\textbf{\color{blue}{46.87}}\\ /\textbf{\color{red}{0.9695}}\end{tabular} & \begin{tabular}[c]{@{}c@{}}47.55\\ /0.9702\end{tabular} \\ \hline \hline
\multirow{3}[8]{*}{Real}
&\multirow{3}[8]{*}{FMD}                                                                  & CF FISH        & \begin{tabular}[c]{@{}c@{}}31.31\\ /\textbf{\color{red}{0.8920}}\end{tabular} & \begin{tabular}[c]{@{}c@{}}31.92\\ /0.8831\end{tabular} & \begin{tabular}[c]{@{}c@{}}22.86\\ /0.4479\end{tabular} & \begin{tabular}[c]{@{}c@{}}32.17\\ /0.8841\end{tabular} & \begin{tabular}[c]{@{}c@{}}\textbf{\color{red}{32.22}}\\ /\textbf{\color{blue}{0.8853}}\end{tabular} & \begin{tabular}[c]{@{}c@{}}32.98\\ /0.9111\end{tabular} \\ \cline{3-9} 
                                                                        &              & CF MICE        & \begin{tabular}[c]{@{}c@{}}37.20\\ /0.9617\end{tabular} & \begin{tabular}[c]{@{}c@{}}37.54\\ /0.9611\end{tabular} & \begin{tabular}[c]{@{}c@{}}30.61\\ /0.7305\end{tabular} & \begin{tabular}[c]{@{}c@{}}\textbf{\color{blue}{38.31}}\\ /\textbf{\color{blue}{0.9634}}\end{tabular} & \begin{tabular}[c]{@{}c@{}}\textbf{\color{red}{38.32}}\\ /\textbf{\color{red}{0.9637}}\end{tabular} & \begin{tabular}[c]{@{}c@{}}38.95\\ /0.9669\end{tabular} \\ \cline{3-9} 
                                                                        &              & TP MICE        & \begin{tabular}[c]{@{}c@{}}33.76\\ /\textbf{\color{red}{0.9157}}\end{tabular} & \begin{tabular}[c]{@{}c@{}}33.34\\ /0.9080\end{tabular} & \begin{tabular}[c]{@{}c@{}}26.24\\ /0.4239\end{tabular} & \begin{tabular}[c]{@{}c@{}}33.89\\ /0.9023\end{tabular} & \begin{tabular}[c]{@{}c@{}}\textbf{\color{red}{33.95}}\\ /\textbf{\color{blue}{0.9084}}\end{tabular} & \begin{tabular}[c]{@{}c@{}}34.40\\ /0.9217\end{tabular} \\ \hline
\multicolumn{3}{|c||}{Inference time}                                                                   & 5.13s                                                   & \textbf{\color{red}{0.06s}}                                                   & 0.99s                                                   & 2.00s                                                      & \multicolumn{2}{c|}{\textbf{\color{blue}{0.21s}}}                                                                                        \\ \hline
\end{tabular}}
    \vspace{-.1in}
\label{table:result_synthetic}
\end{table}

\noindent\textbf{Real-world noise}  \ \
The results on \textbf{FMD} dataset are reported in the bottom of Table \ref{table:result_synthetic}. 
From the table, we observe again that FBI-D consistently outperforms other baselines.
Moreover, D-BSN suffers from serious performance degradation, and this illustrates another example of poor applicability of D-BSN when the noise is strong.

\textcolor{\tsmcolor}{Table \ref{table:result_SIDD/DND} shows the results on \textbf{SIDD} \cite{(SIDD)SIDD_2018_CVPR} and \textbf{DND} \cite{(DND)plotz2017benchmarking} datasets. 
In DND, since the  $50$ noisy test images are available, we regard this set as both training and test set. 
Note this is perfectly possible since all methods in Table \ref{table:result_SIDD/DND} do not require any clean images for training. 
In addition, particularly for the DND results, we also 
report the results of BP-AIDE and FBI-D, which are trained on the training set of SIDD.
For clarity, we present two versions of BP-AIDE and FBI-D trained on different datasets; ``(SIDD)'' and ``(DND)'' stand for the dataset used for training, respectively. 
From the table, firstly, we reconfirm the similar tendency as Table \ref{table:result_synthetic}; \textit{i.e.}, FBI-D mostly dominates other baselines and enjoys fast inference time. 
Note SIDD and DND datasets contain a mixture of various noise levels, hence, this result shows the robustness and effectiveness of our FBI-D for the real ``mismatched'' noise case.
Secondly, from the strong performance of ``FBI-D (SIDD)'' for both SIDD and DND, we verify the strong generalization capability of FBI-D. We also note that, as we show in the S.M., the performance of ``FBI-D (SIDD)'' is also competitive with supervised trained models. 
Finally, the visualization results in Figure \ref{fig:visualization} re-emphasize the strength of ``FBI-D (SIDD)''; it reconstructs the detailed texture much better than any other baselines.}

\begin{table}[h!]
\vspace{-.05in}
\caption{PSNR(dB)/SSIM on  SIDD and DND dataset. The colored texts are as before.
}
\centering
\smallskip\noindent
\resizebox{.98\linewidth}{!}{
\begin{tabular}{|c|c|c||c|c|c|c|c|c|c|}
\hline
\multicolumn{3}{|c||}{Dataset} & GAT+BM3D                                                 & N2V                                                     & D-BSN                                                   & \begin{tabular}[c]{@{}c@{}}BP-AIDE\\ (SIDD)\end{tabular} & \begin{tabular}[c]{@{}c@{}}BP-AIDE\\ (DND)\end{tabular} & \begin{tabular}[c]{@{}c@{}}FBI-D\\ (SIDD)\end{tabular}    & \begin{tabular}[c]{@{}c@{}}FBI-D\\ (DND)\end{tabular}     \\ \hline \hline
\multirow{4}[10]{*}{Real} 
&\multirow{2}[4]{*}{SIDD}  & RAW  & \begin{tabular}[c]{@{}c@{}}48.52\\ /\textbf{\color{blue}{0.9800}}\end{tabular}  & \begin{tabular}[c]{@{}c@{}}46.30\\ /0.9760\end{tabular} & \begin{tabular}[c]{@{}c@{}}37.16\\ /0.8390\end{tabular} & \begin{tabular}[c]{@{}c@{}}\textbf{\color{blue}{50.45}}\\ /\textbf{\color{red}{0.9900}}\end{tabular}  & -                                                       & \begin{tabular}[c]{@{}c@{}}\textbf{\color{red}{50.57}}\\ /\textbf{\color{red}{0.9900}}\end{tabular} & -                                                       \\ \cline{3-10} 
    &                       & sRGB & \begin{tabular}[c]{@{}c@{}}34.61\\ /\textbf{\color{blue}{0.8789}}\end{tabular} & \begin{tabular}[c]{@{}c@{}}32.85\\ /0.8470\end{tabular} & \begin{tabular}[c]{@{}c@{}}24.07\\ /0.4999\end{tabular} & \begin{tabular}[c]{@{}c@{}}37.91\\ /\textbf{\color{red}{0.9420}}\end{tabular}  & -                                                       & \begin{tabular}[c]{@{}c@{}}\textbf{\color{red}{38.07}}\\ /\textbf{\color{red}{0.9420}}\end{tabular} & -                                                       \\  \cline{2-10} 
&\multirow{2}[6]{*}{DND}   & RAW  & \begin{tabular}[c]{@{}c@{}}47.53\\ /0.9761\end{tabular}  & \begin{tabular}[c]{@{}c@{}}45.41\\ /0.9688\end{tabular} & \begin{tabular}[c]{@{}c@{}}39.63\\ /0.8642\end{tabular} & \begin{tabular}[c]{@{}c@{}}\textbf{\color{blue}{47.75}}\\ /\textbf{\color{blue}{0.9770}}\end{tabular}  & \begin{tabular}[c]{@{}c@{}}47.60\\ /0.9732\end{tabular} & \begin{tabular}[c]{@{}c@{}}\textbf{\color{red}{48.02}}\\ /\textbf{\color{red}{0.9787}}\end{tabular} & \begin{tabular}[c]{@{}c@{}}47.53\\ /0.9706\end{tabular} \\ \cline{3-10} 
              &         & sRGB & \begin{tabular}[c]{@{}c@{}}37.98\\ /0.9203\end{tabular}  & \begin{tabular}[c]{@{}c@{}}35.82\\ /0.9022\end{tabular} & \begin{tabular}[c]{@{}c@{}}30.23\\ /0.7095\end{tabular} & \begin{tabular}[c]{@{}c@{}}\textbf{\color{blue}{38.79}}\\ /\textbf{\color{blue}{0.9446}}\end{tabular}  & \begin{tabular}[c]{@{}c@{}}38.60\\ /0.9259\end{tabular} & \begin{tabular}[c]{@{}c@{}}\textbf{\color{red}{38.98}}\\ /\textbf{\color{red}{0.9451}}\end{tabular} & \begin{tabular}[c]{@{}c@{}}38.56\\ /0.9185\end{tabular} \\ \hline
              \multicolumn{3}{|c||}{Inference time}                                                                   & 5.13s                                                   & \textbf{\color{red}{0.06s}}                                                   & 0.99s                                                   & \multicolumn{2}{c|}{2.00s}                                                      & \multicolumn{2}{c|}{\textbf{\color{blue}{0.21s}}}                                                                                        \\ \hline
\end{tabular}
}
    \vspace{-.2in}
\label{table:result_SIDD/DND}
\end{table}

\subsection{Ablation study}
\noindent\textbf{Ablation study on PGE-Net} \ \
\textcolor{\tsmcolor}{Here, we analyze the effect of the loss function of PGE-Net (\ref{eq:loss_noise_est}) by comparing with the \textit{naive} supervised estimation model. 
Table \ref{table:ablation_estimation} shows the PSNR/SSIM values of GAT+BM3D using estimated noise parameters from different methods on Fivek dataset corrupted with a mixture of noise levels ($\alpha \in [0,0.16^2], \sigma \in [0,0.06]$).
For comparison, we trained ``Sup (MSE)''  with exact same architecture as PGE-Net by minimizing the MSE between $(\alpha, \sigma)$ and $(\hat{\alpha},\hat{\sigma})$.
The result of PGE-Net overwhelming that of ``Sup (MSE)'' may seem counter-intuitive, but we verify that when GAT is done with the estimated noise parameters of PGE-Net and ``Sup (MSE)'', the average variances of the noise in the transformed images become $1.03$ and $1.58$, respectively. Thus, it turns out that ``Sup (MSE)'' merely focuses on estimating $\alpha$ and $\sigma$ via minimizing the squared error, but its estimated parameters do not necessarily result in stabilized variance after carrying out GAT using them. We believe this result again shows the effectiveness and strength of our loss function (\ref{eq:loss_noise_est}) for PGE-Net.}
\begin{table}[h!]
\caption{Ablation studies of PGE-Net on Fivek (with a mixed noise)
}
\centering
\smallskip\noindent
\resizebox{0.9\linewidth}{!}{
\begin{tabular}{|c||c|c|c|c|}
\hline
PSNR/SSIM         & Foi \cite{(Foi)foi2008practical}           & Liu \cite{(pg_Liu)liu2014practical}           & \textbf{PGE-Net}        & Sup (MSE)       \\ \hline \hline
GAT+BM3D
& 37.23 / 0.9113 & 38.58 / 0.9336 & \textbf{38.72} / \textbf{0.9350} & 35.84 / 0.8674 \\ \hline
\end{tabular}
}
\label{table:ablation_estimation}
\end{table}

\noindent\textbf{Ablation study on FBI-Net} \ \
We demonstrate the necessity of each component of FBI-Net. Table \ref{table:ablation_model} compares the PSNR/SSIM values of the networks with and without each component on Fivek ($\alpha = 0.01, \sigma = 0.02$). 
RC and RM denotes Residual Connection and Residual Module, respectively. 
All networks are trained by a supervised way.
 From the table, we observe a serious performance degradation when any component of FBI-Net is absent, hence, conclude that RC and RM in our FBI-Net are essential for the successful training of FBI-D. 
\begin{table}[h!]
\vspace{-.1in}
\caption{Ablation studies of FBI-Net on Fivek ($\alpha = 0.01, \sigma = 0.02$).}
\centering
\smallskip\noindent
\resizebox{.98\linewidth}{!}{
\begin{tabular}{|c||c|c|c|c|c|c|c|c|}
\hline
Component                                            & \textbf{FBI-Net}                                                   & Case1                                                   & Case2                                                   & Case3                                                   & Case4                                                   & Case5                                                   & Case6                                                   & Case7                                                   \\ \hline \hline
Outer RC                                             & \textbf{\cmark}                                                       & \xmark                                                       &  \cmark                                                      & \cmark                                                       & \xmark                                                       & \xmark                                                       & \cmark                                                       & \xmark                                                       \\ \hline
Inner RC                                             & \textbf{\cmark}                                                       & \cmark                                                       & \xmark                                                       & \cmark                                                       & \xmark                                                       & \cmark                                                       & \xmark                                                       & \xmark                                                       \\ \hline
RM                                                   & \textbf{\cmark}                                                       & \cmark                                                       & \cmark                                                       & \xmark                                                       & \cmark                                                       & \xmark                                                       & \xmark                                                       & \xmark                                                       \\ \hline
\begin{tabular}[c]{@{}c@{}}PSNR\\ /SSIM\end{tabular} & \textbf{\begin{tabular}[c]{@{}c@{}}44.38\\ /0.9537\end{tabular}} & \begin{tabular}[c]{@{}c@{}}44.11\\ /0.9506\end{tabular} & \begin{tabular}[c]{@{}c@{}}44.31\\ /0.9529\end{tabular} & \begin{tabular}[c]{@{}c@{}}44.19\\ /0.9520\end{tabular} & \begin{tabular}[c]{@{}c@{}}44.24\\ /0.9525\end{tabular} & \begin{tabular}[c]{@{}c@{}}28.82\\ /0.5103\end{tabular} & \begin{tabular}[c]{@{}c@{}}28.84\\ /0.5118\end{tabular} & \begin{tabular}[c]{@{}c@{}}28.82\\ /0.5103\end{tabular} \\ \hline
\end{tabular}}
    \vspace{-.2in}
\label{table:ablation_model}
\end{table}

\section{\textcolor{\tsmcolor}{Concluding Remarks}}
We proposed FBI-Denoiser which resolves the computational complexity issue of BP-AIDE by devising PGE-Net, which is much faster than conventional Poisson-Gaussian noise estimation ($\times 2000$), and FBI-Net, which is an efficient blind spot network. \textcolor{\tsmcolor}{We showed FBI-Denoiser achieves the state-of-the-art blind image denoising performance solely based on ``single'' noisy images with much faster inference time on various synthetic/real noise benchmark datasets.}
\section*{\textcolor{\tsmcolor}{Acknowledgment}}
This work was supported in part by NRF Mid-Career Research Program [NRF-2021R1A2C2007884] and IITP grant [No.2019-
0-01396, Development of framework for analyzing, detecting, mitigating of bias in AI model and
training data], funded by the Korean government (MSIT).

\newpage
{\small
\bibliographystyle{ieee_fullname}
\bibliography{egbib}
}
\clearpage


\section*{Supplementary material}

# Supplementary Material for
# FBI-Denoiser: Fast Blind Image Denoiser for
# Poisson-Gaussian Noise

Jaeseok Byun[1]*, Sungmin Cha[1]*, and Taesup Moon[2]†
[1]Department of ECE, Sungkyunkwan University, Suwon, Korea
[2]Department of ECE, Seoul National University, Seoul, Korea
{wotjr3868, csm9593}@skku.edu, tsmoon@snu.ac.kr

## 1. Implementation details

### 1.1. Software Platform / hyperparameters for training

All experiments were conducted by eight-core Intel(R) Xeon(R) Silver 4110 CPU @ 2.10GHz, 128GB RAM and an NVIDIA Titan XP. Learning rages of 0.001 and 0.0001 were used for training FBI-Net and PGE-Net, respectively. We used Adam [7] optimizer and drop the learning rate by half every ten epochs. Additionally, we used Pytorch [9] with CUDA 10.2 to implement FBI-Net and PGE-Net, with mini-batch size as 1. For the Gaussian noise level estimation [4], the patch size is set to eight.

### 1.2. Gaussian noise estimation [4] (Tensorized)

[4] is a PCA-based Gaussian noise level estimation algorithm, developed with an intuition that image patches generated from a clean image often lie in a low-dimensional subspace. Thus, their core idea is to find redundant dimensions of a noisy image for capturing noise component by carefully eliminating the principal dimensions of a image. Namely, they start removing eigenvalues of a covariance matrix of noisy patches from the largest value, and stop at the right time.

As described in [4, Algorithm 1], it is mainly composed of three steps. First, decompose a single noisy image $\boldsymbol{Y} \in \mathbb{R}^{N \times N}$ to generate vectorized patches with size $d^2$. In second step, calculate the eigenvalues $\boldsymbol{S} = \{\lambda_i\}_{i=1}^{d^2}$ of the covariance matrix of patches. If eigenvalues are sorted in ascending order, *i.e.*, $\lambda_1 \leq \lambda_2 \leq \cdots \leq \lambda_{d^2}$, $\boldsymbol{S}$ can be represented as $\boldsymbol{S}_n \cup \boldsymbol{S}_p$ where $\boldsymbol{S}_n = \{\lambda_i\}_{i=1}^{m}$ indicates a set of eigenvalues for redundant dimensions and $\boldsymbol{S}_p = \{\lambda_i\}_{i=m+1}^{d^2}$ indicates a set of eigenvalues for the principal dimensions. Then, the last step of [4] is carefully finding a point that divides $\boldsymbol{S}_n$ and $\boldsymbol{S}_p$, and then calculating the noise variance with $\boldsymbol{S}_n$. For this step, [4] proposed effective *iterative* dimension selection procedure that stops iterations where mean and median of subset $\boldsymbol{S}_n$ are equal. They theoretically verified that this stopping point is a right point to separate $\boldsymbol{S}_n$ and $\boldsymbol{S}_p$. More details about algorithm and theoretical justifications can be found in [4, Section 2]. The original implementation is initialized with $\boldsymbol{S}_p = \varnothing$ and $\boldsymbol{S}_n = \boldsymbol{S}$. By denoting the mean and median value of the subset $\boldsymbol{S}_n$ as $\gamma$ and $\psi$ respectively, the difference between $\gamma$ and $\psi$ is iteratively calculated for

---

*Equal contribution.
†Corresponding author (E-mail: tsmoon@snu.ac.kr)

finding outliers in the subset $S_n$. If $\gamma \neq \psi$, take out the largest value in the subset $S_n$ and put it into the subset $S_p$. This procedure is iterated until $\gamma = \psi$. Then, the final result of noise estimation is obtained by $\sqrt{\gamma}$.

---

**Algorithm 1** Gaussian Noise estimation [4] (Tensorized)

---

**Require:** A noisy image $Y \in \mathbb{R}^{N \times N}$, Patch size $d$

    ▷ /* Step1: Decompose image */

1: Decompose a noisy image $Y$ into vectorized multiple patches $\{y_t\}_{t=1}^{s}$ with size $d^2$ where $s = (N - d + 1)^2$

2: $u \leftarrow \frac{1}{s} \sum_{t=1}^{s}(y_t)$, $\Sigma_y \leftarrow \frac{1}{s} \sum_{t=1}^{s}(y_t - u)(y_t - u)^T$

    ▷ /* Step2: Calculate the eigenvalues */

3: Calculate eigenvalues $\lambda_i$ of $\Sigma_y$ with ascending order $\lambda_1 \leq \lambda_2 \leq \cdots \leq \lambda_{d^2}$

4: $\Lambda \in \mathbb{R}^{d^2 \times d^2} \leftarrow$ a diagonal matrix where $\Lambda_{ii} = \lambda_i$

    ▷ /* Step3: Finding $\sigma$ without iterative procedure */
    /* Generate two lower triangular matrices */

5: $A_\Lambda \leftarrow A_1 \cdot \Lambda$ (dot product)

6: Calculate a mean vector $\tau \in \mathbb{R}^{d^2}$
where $\tau_i = \frac{1}{k_i} \sum_{j=1}^{d^2} (A_\Lambda)_{ij}$ and $k_i$ is the number of nonzero elements in $(A_\Lambda)_i$

7: $A_\tau \leftarrow \text{Diag}(\tau) \cdot A_1$ (dot product)

    /* Identifying the median value with the masking scheme */

8: Generate masking matrix $M_{big} \in \mathbb{R}^{d^2 \times d^2}$ and $M_{small} \in \mathbb{R}^{d^2 \times d^2}$
where $(M_{big})_{ij} = \mathbb{1}\{(A_\Lambda)_{ij} > (A_\tau)_{ij}\}$ and $(M_{small})_{ij} = \mathbb{1}\{(A_\Lambda)_{ij} < (A_\tau)_{ij}\}$

9: Generate counting vectors $c_{big} \in \mathbb{R}^{d^2}$ and $c_{small} \in \mathbb{R}^{d^2}$
where $(c_{big})_i = \sum_{j=1}^{d^2} (M_{big})_{ij}$ and $(c_{small})_i = \sum_{j=1}^{d^2} (M_{small})_{ij}$

10: Generate a masking vector $m \in \mathbb{R}^{d^2}$ where $m_i = \mathbb{1}\{(c_{big})_i == (c_{small})_i\}$.

11: $\sigma \leftarrow \sqrt{\max(m \odot \tau)}$

12: **return** $\sigma$

---

As we mentioned in Section 4.1 (manuscript), we replaced this *iterative* procedure in [4] with tensor operations. It is straight-forward to change operations of first and second step to tensor operations, but the last step is not. Thus, we mainly describe the last step. Before describing the tensorized version of [4], we introduce a few more notations. Let's define a lower triangular matrix as $A \in \mathbb{R}^{d^2 \times d^2}$ and a specific lower triangular matrix which has 1 as all nonzero elements as $A_1 \in \mathbb{R}^{d^2 \times d^2}$. Moreover, define $M \in \mathbb{R}^{d^2 \times d^2}$ as masking matrix and $m \in \mathbb{R}^{d^2}$ as masking vector such that $M_{ij} \in \{0, 1\}$ and $m_i \in \{0, 1\}$. $\text{Diag}(a)$ is the $d^2 \times d^2$ diagonal matrix whose diagonal elements are the elements of the vector $a \in \mathbb{R}^{d^2}$ and $\odot$ denotes element-wise multiplication.

As can be seen in Algorithm 1, instead of *iterative* procedure for calculating the mean value $\gamma$ and median value $\psi$, we generate two lower triangular matrices which can compute all possible mean values and identify whether those mean values are median or not *at once*. For identifying each mean value is a median or not, every possible mean values should be calculated and compared with eigenvalues. Namely, the individual averages of the eigenvalues of each candidate of $S_n$ are computed first, and then each mean value is compared with the eigenvalues of that candidate. Thus, firstly, we make a low triangular matrix $A_\Lambda$ which consists of candidates for $S_n$. For this, $A_\Lambda$ is calculated by the dot product of $A_1$ and eigenvalue matrix $\Lambda$. Considering only the non-zero element, the first row represents the smallest $S_n$ containing only $\lambda_1$, and the last row represents the largest $S_n$, *i.e*, $S$. Secondly, we calculate a mean vector $\tau$ and generate a low triangular matrix $A_\tau$ for comparing with $A_\Lambda$. By using $A_\Lambda$, a mean vector $\tau$ consisting of every possible mean values is calculated. Then, $A_\tau$ is calculated by the dot product of $Diag(\tau)$ and $A_1$. Note that the non-zero elements in each row of $A_\tau$ all have the same value, and the single non-zero element in i-th row of $A_\tau$ is the average value of the i-th row of $A_\tau$. Namely, the non-zero elements of the i-th row of $A_\tau$ are composed of i-repeats of the average of i-th row of $A_\tau$.

Now, all possible mean values can be compared with corresponding eigenvalues at once. To identify the median value, two masking matrices $M_{big}$ and $M_{small}$ are generated, derived from conditions that the inequality for the above two lower triangular matrices are opposite. By comparing two counting vectors, the final masking vector $m$ is calculated. Since we assume the eigenvalues are sorted with ascending order, the maximum value of $m \odot \tau$ is the final estimates for the noise variance.

## 2. Detailed Experimental Results

### 2.1. Toy example for verifying the Section 5.2 (manuscript)

As we discussed in Section 5.2 (manuscript), whereas the accuracy of PGE-Net for estimating $\sigma$ is not high, the GAT+BM3D or BP-AIDE framework with the GAT-transformed images (using underestimated $\hat{\sigma}$ of PGE-Net) still shows great blind denoising performance. This somewhat counter-intuitive can be explained by our empirical verification; We observed that the underestimation of PGE-Net for $\sigma$ is not a major issue in achieving that GAT-transformed image (with the estimated parameters) has the stabilized noise variance close to 1. To verify this, consider the simple toy example shown in Figure 1. We generated four different homogeneous image patches of size $256 \times 256$ with four different pixel values $(0.2, 0.7, 0.8, 0.3)$, respectively, and corrupted them with a Poisson-Gaussian noise ($\alpha = 0.1$, $\sigma = 0.02$) (Fig. 1a). Note that due to the source-dependence of the noise, the four image patches have different noise levels as shown in Fig. 1b. Now, Fig. 1c shows the GAT-transformed images with the estimated noise parameters from the PGE-Net (($\hat{\alpha}, \hat{\sigma}$) for each image is given in Fig.1c), and Fig. 1d shows the corresponding noise of the GAT-transformed images (obtained by subtracting the mean of each image). Notice that while $\hat{\sigma}$ of PGE-Net is indeed much smaller than the true $\sigma$, as in Table 3 (manuscript), Fig. 1d shows that the source-dependent noise is almost removed after GAT, and the variance of the remaining noise gets *very close* to 1. Note PGE-Net is trained *exactly* with this objective (Eq.(8) in manuscript), and we believe this example clearly shows why running the BP-AIDE framework with the GAT-transformed images (using underestimated $\hat{\sigma}$ of PGE-Net) still shows great blind denoising performance.

### 2.2. Comparison with supervised trained methods in SIDD / DND

In this section, we additionally compare the performance of FBI-Denoiser with supervised trained models such as UPI [3] and CycleISP [11]. As we mentioned in Introduction (manuscript), CycleISP and UPI modeled in-camera processing pipeline (ISP) and achieved the state-of-the-art performance in

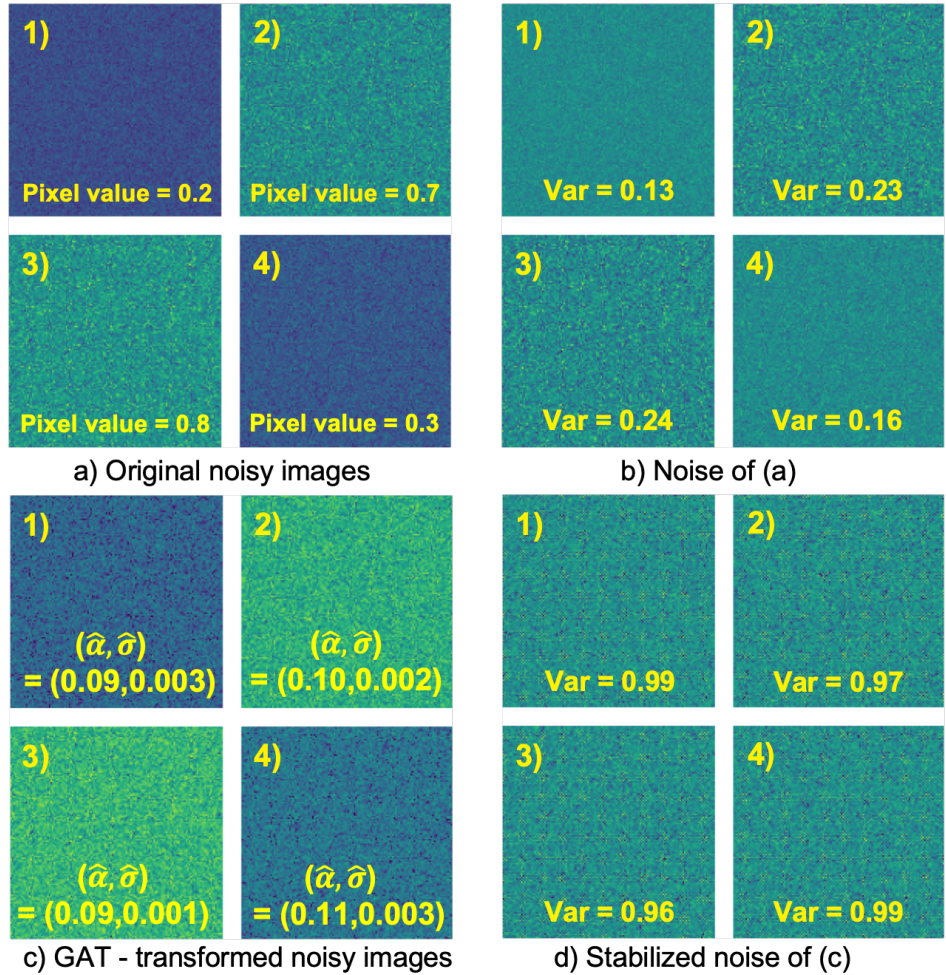
Figure 1. A toy example for Section 5.2 (manuscript)

SIDD and DND dataset by generating noisy and clean pairs of raw-RGB or sRGB images from clean synthetic sRGB images. Their approaches are trained in a supervised way with pairs of generated pairs. Moreover, they used an additional input channel which is composed of per-pixel estimates of the noise standard-deviation for a noisy image. Note the information about Poisson-Gaussian noise parameters is required for getting these per-pixel estimates. However, they assumed that the information about noise parameters of test noisy images is given in both training and evaluation phase which is unrealistic in the real-world application. In evaluation, they used true noise parameters which are given by DND and SIDD websites [1, 2]. Moreover, in training, as can be seen in [3, Section 3.1], they specify the sampling range of noise parameters $(\alpha, \sigma)$ based on the maximum and minimum values of the noise parameters of test noisy images, and further define more specific sampling ranges by performing regression fitting with the noise parameters of the test noisy images. Namely, they used the additional information about noise characteristics in both training and evaluation phase.

Thus, we recalculated their results in more realistic settings which the information about noise characteristics is not given in training or evaluation. Additionally, we measured the performance of

Table 1. PSNR(dB)/SSIM on SIDD and DND dataset. Red and blue denote the highest and second highest results, respectively, among models except supervised trained models.

| Type | Algorithm | SIDD | | | | DND | | | |
|---|---|---|---|---|---|---|---|---|---|
| | | Raw | | sRGB | | Raw | | sRGB | |
| | | PSNR | SSIM | PSNR | SSIM | PSNR | SSIM | PSNR | SSIM |
| unsup | GAT+BM3D | 48.52 | 0.9800 | 34.61 | 0.8789 | 47.53 | 0.9761 | 37.98 | 0.9203 |
| | N2V | 46.30 | 0.9760 | 32.85 | 0.8470 | 45.41 | 0.9688 | 35.82 | 0.9022 |
| | D-BSN | 37.16 | 0.8390 | 24.07 | 0.4999 | 39.63 | 0.8642 | 30.23 | 0.7095 |
| | BP-AIDE (DND) | - | - | - | - | 47.60 | 0.9732 | 38.60 | 0.9259 |
| | FBI-D (DND) | - | - | - | - | 47.53 | 0.9706 | 38.56 | 0.9185 |
| | BP-AIDE (SIDD) | 50.45 | 0.9900 | 37.91 | 0.9420 | 47.75 | 0.9770 | 38.79 | 0.9446 |
| | FBI-D (SIDD) | 50.57 | 0.9900 | 38.07 | 0.9420 | 48.02 | 0.9787 | 38.98 | 0.9451 |
| sup | UPI (25k, blind) with Liu [8] | 47.89 | 0.9860 | 36.59 | 0.9280 | 46.96 | 0.9756 | 38.34 | 0.9450 |
| | UPI (25k) with Liu [8] | 48.95 | 0.9870 | 37.15 | 0.9330 | 47.58 | 0.9756 | 38.92 | 0.9413 |
| | MCU-Net | 48.80 | 0.9900 | 36.54 | 0.8750 | - | - | - | - |
| | UPI (1M) | - | - | - | - | 48.89 | 0.9820 | 40.17 | 0.9620 |
| | CycleISP (1M) | 52.41 | 0.9930 | 39.47 | 0.9180 | 49.13 | 0.9830 | 40.50 | 0.9660 |

them when the number of training pairs is limited to 25000 (originally, 1 million) which is the same number with training patches in FBI-Denoiser. Note that the training code of CycleISP is not available, we only measured the performance of UPI in realistic settings, thus, we mainly analyzed the results of UPI. Since the main difference between CycleISP and UPI lies in the detailed implementation of data conversion step from sRGB to raw-RGB, we believe that results of CycleISP which is trained in a realistic setting will have a similar tendency to the remeasured results of UPI.

The first realistic setting "UPI (25k) with Liu [8]" denotes the result of UPI which is trained with the original sampling procedure of UPI, but evaluated with estimated noise parameters from Liu [8]. The more realistic setting "UPI (25k, blind) with Liu [8]" denotes the result of UPI which is trained with a blind uniform sampling procedure ($\alpha \in [0, 0.16^2], \sigma \in [0, 0.06]$) as in [6], and evaluated with estimated noise parameters from Liu [8]. Note that the inference time for noise estimation is additionally needed for UPI in these realistic settings.

The results of MCU-Net, UPI (1M), and CycleISP were taken from their original papers [?, 3, 11] since their training takes too long. Note the result of UPI which is trained with 1 million noisy images in SIDD is not reported since [3] only reports the DND evaluation results. As can be seen in Table 1, in realistic settings, the performance of UPI is significantly degraded. As a result, FBI-Denoiser outperforms not only unsupervised trained methods but also supervised trained methods including UPI (in realistic settings) and MCU-Net. Note that MCU-Net is also a strong baseline that ranked high in the NTIRE 2020 Challenge on Real Image Denoising - Track1: rawRGB. These results imply that FBI-Denoiser which is trained solely based on single noisy images is enough to compete with supervised trained methods, and given sufficient noisy images, the performance of FBI-Denoiser could be further improved.

## 2.3. Experiments on the weak noise cases

Although we have shown that our FBI-Denoiser obtains the state-of-the-art denoising performance in real-world noisy benchmark datasets which contain a mixture of various noise levels, the performance of FBI-Denoiser is relatively low in weak noise cases as can be seen (0.01, 0.0002) case of Fivek dataset in Table 5 (manuscript) and this performance degradation becomes severe as the signal to noise ratio (SNR) increases. Note this performance degradation is caused by the original loss function (Eq. 7 in manuscript) of BP-AIDE, not the PGE-Net and FBI-net we proposed. As we described in Section 4.2 (manuscript), we followed the same procedure of BP-AIDE which uses the loss function (Eq. 7 in manuscript) for training a denoiser. FC-AIDE and BP-AIDE achieve higher

performance than N2V by using additional information, the noisy pixel $Z_i$, as can be seen in their pixelwise affine denoiser and loss function (Eq. 5,7 in manuscript), but, it's not always good to use whole $Z_i$. If the noise intensity is too strong, it may be better not to use $Z_i$ for training since $Z_i$ is are too noisy. Thus, we believe that the process which decides how much $Z_i$ will be utilized based on the noise intensity would be needed to get better performance. But, BP-AIDE and our FBI-Denoiser set the range of slope coefficients $a_1(w; \cdot)$ as a single fixed range for all noisy images. This range is set between 0 and a specific maximum value as can be seen in Section 5.1 (manuscript). Although they show great performance in real-world noise benchmarks with only this fixed range of $a_1(w; \cdot)$, their performance is limited in some noise cases due to the fixed range of slope coefficients $a_1(w; \cdot)$. Thus, if a loss function that adaptively designates the range of slope coefficient $a_1(w; \cdot)$ according to noise intensity is devised, we expect that overall performance in real-world noisy benchmark datasets would be further enhanced. We left this as our future work.

### 2.4. Decision rules for the failure cases of Liu [8] and Foi [5] in BP-AIDE

As mentioned in Section 5.2 (manuscript), BP-AIDE has to set up rules for handling the extremely small values of $\hat{\alpha}$ which are the failure cases of Liu [8] and Foi [5]. Two decision rules were suggested in their source code; First, if $\hat{\alpha} < 1E - 7$, set $\hat{\alpha}$ as 1 in both training and evaluation. Second, if $\hat{\alpha} < 1E - 7$, exclude its noisy image from the training. The best results among them are reported in all experiments. Note that the performance deviation of BP-AIDE is relatively high due to the estimation failure and its handling.

### 2.5. Inference time of noise estimation

We additionally measured the inference time according to the different size of patches. As seen in Table 2, the inference time of Foi [5] and Liu [8] increases as the patch size increases. In particular, the inference time of Liu [8] drastically increased when the patch size is doubled. On the other hand, PGE-Net keeps the inference time almost constant. Thus, we verified that the efficiency of PGE-Net outperforms other algorithms in all various sized patches.

Table 2. Experimental results of inference time of noise estimation

| Dataset | Patch size | Foi [5] | Liu [8] | PGE-Net |
|---------|------------|---------|---------|---------|
|         | 256x256    | 0.69    | 0.72    | 0.0017  |
| Fivek   | 512x512    | 2.5     | 1.8     | 0.0020  |
|         | 1024x1024  | 2.83    | 6.88    | 0.0023  |

### 2.6. Network architecture of PGE-Net

Table 3 compares the results of GAT+BM3D using estimated noise parameters from two PGE-Nets which have the different number of layers. The experimental setting is same as ablation studies on PGE-Net in Section 5.4 (manuscript), which uses the uniformly sampled noise parameters ($\alpha \in [0, 0.16^2], \sigma \in [0, 0.06]$). As can be seen in Table 3, the difference in performance between 3-layer and 1-layer is *very* small. Thus, it is highly likely that there will be a simpler and more efficient network architecture than PGE-Net, which simply adopted three layer U-Net, but we did not attempt to find the optimized architecture of PGE-Net and left it as our future work.

Table 3. Network architecture of PGE-Net

| Algorithms | PGE-Net (3 layer) | PGE-Net (1 layer) |
|------------|-------------------|-------------------|
| GAT+BM3D PSNR/SSIM | **38.72 / 0.9350** | 38.61 / 0.9340 |

## 2.7. Visualization Results on DND

We visualized an additional benchmark image in DND [10]. We clearly observe that FBI-Denoiser and BP-AIDE achieve a superior performance in terms of both PSNR and SSIM compared with other baselines. However, note FBI-Denoiser restored the detailed texture better than BP-AIDE with a faster reference time.

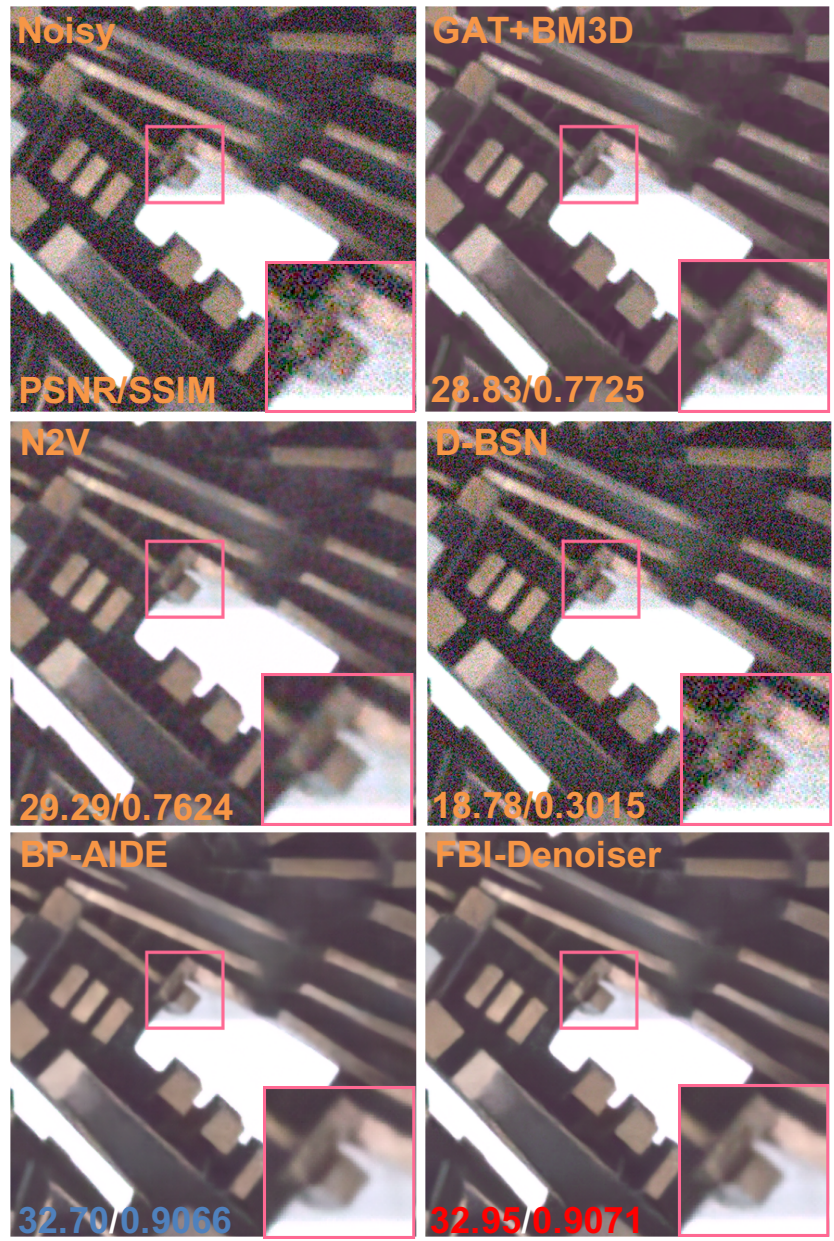

Figure 2. Visualization results of DND



\end{document}